\documentclass[3p,times]{article}

\usepackage{amsmath}
\usepackage{url}
\usepackage{graphicx}
\usepackage{subfigure}
\usepackage{xspace}
\usepackage{hyperref}

\usepackage[utf8]{inputenc}

\usepackage{listings}
\usepackage{capt-of}

\usepackage{balance}
\usepackage{multicol}
\usepackage{caption}

\usepackage{color}

\usepackage{wrapfig}

\usepackage{paralist}

\definecolor{blue}{rgb}{0,0,1}
\definecolor{violet}{rgb}{0.5,0,0.5}

\definecolor{darkred}{rgb}{0.5,0,0}
\definecolor{darkblue}{rgb}{0,0,0.75}
\definecolor{darkgreen}{rgb}{0,0.5,0}

\usepackage[affil-it]{authblk}

\usepackage{geometry}

\hypersetup{%
  colorlinks=false,%
  pdfborder = 0 0 0,%
  pdfproducer={},%
  bookmarksnumbered,%
  bookmarksopen=false,%
  pdfstartview=Fit,%
  linkcolor=black,%
  citecolor=black,%
  filecolor=black,%
  menucolor=black,%
  urlcolor=black,%
  plainpages=false,%
  hypertexnames=false%
}

\newcommand{\doublefig}[8]{%
 \begin{figure*}%
  \centering%
  \subfigure[#1]{\includegraphics[width=1\columnwidth]{#2}\label{#3}}%
  \hfill%
  \subfigure[#4]{\includegraphics[width=1\columnwidth]{#5}\label{#6}}%
  \caption{#7}\label{#8}%
 \end{figure*}%
}

\newcommand{\singlefig}[3]{%
 \begin{figure}[htb]%
  \centering%
  \includegraphics[width=1\columnwidth]{#1}%
  \captionof{figure}{#2}%
  \label{#3}%
 \end{figure}%
}

\newcommand{\caf}{CAF\xspace}

\newcommand{\C}{C\xspace}
\newcommand{\cpp}{C++\xspace}

\newcommand{\opencl}{OpenCL\xspace}
\newcommand{\openclc}{OpenCL\,C\xspace}

\definecolor{red}{rgb}{1,0,0}
\definecolor{blue}{rgb}{0,0,1}
\definecolor{grey}{rgb}{0.5,0.5,0.5}
\definecolor{violet}{rgb}{0.5,0,0.5}
\definecolor{midblue}{rgb}{0,0,0.75}
\definecolor{darkred}{rgb}{0.5,0,0}
\definecolor{darkblue}{rgb}{0,0,0.5}
\definecolor{darkgreen}{rgb}{0,0.5,0}
\definecolor{orchid}{rgb}{0.45,0.51,0.9}
\definecolor{midgrey}{rgb}{0.5,0.5,0.5}

\emergencystretch=\maxdimen
\DeclareCaptionFormat{listing}{\rule{\dimexpr\textwidth\relax}{0.4pt}\par\vskip1pt~#1#2#3}
\begin{document}

\widowpenalty10000
\clubpenalty10000

\setlength{\pdfpageheight}{\paperheight}
\setlength{\pdfpagewidth}{\paperwidth}

\title{OpenCL Actors -- Adding Data Parallelism\\ to  Actor-based Programming with CAF}

\author[1]{Raphael Hiesgen\thanks{Corresponding author: raphael.hiesgen@haw-hamburg.de}}

\author[1]{Dominik Charousset}

\author[1]{Thomas C. Schmidt}

\affil[1]{\small{Hamburg University of Applied Sciences

Internet Technologies Group

Department Informatik, HAW Hamburg

Berliner Tor 7, D-20099 Hamburg, Germany}}

\maketitle

\begin{abstract}

The actor model of computation has been designed for a seamless support of concurrency and distribution. However, it remains unspecific about data parallel program flows, while available processing power of modern many core hardware such as graphics processing units (GPUs) or coprocessors increases the relevance of data parallelism for general-purpose computation.

In this work, we introduce OpenCL-enabled actors to the C++ Actor Framework (CAF). This offers a high level interface for accessing any OpenCL device without leaving the actor paradigm. The new type of actor is integrated into the runtime environment of CAF and gives rise to transparent message passing in distributed systems on heterogeneous hardware. Following the actor logic in CAF, OpenCL kernels can be composed while encapsulated in C++ actors, hence operate in a multi-stage fashion on data resident at the GPU. Developers are thus enabled to build complex data parallel programs from primitives without leaving the actor paradigm, nor sacrificing performance. Our evaluations on commodity GPUs, an Nvidia TESLA, and an Intel PHI reveal the expected linear scaling behavior when offloading larger workloads. For sub-second duties, the efficiency of offloading was found to largely differ between devices. Moreover, our findings indicate a negligible overhead over programming with the native OpenCL API. 

\end{abstract}


\lstset{
  language=C++,
  morekeywords={constexpr,nullptr,size\_t,this,\_\_global,\_\_kernel},%
  frame=top,frame=bottom,
  basicstyle=\small\normalfont\ttfamily,
  stepnumber=1,
  numbersep=10pt,
  tabsize=2,
  extendedchars=true,
  breaklines=true,
  captionpos=t,
  numbers=left,
  numberstyle=\small\color{midgrey},
  showspaces=false,
  showtabs=false,
  xleftmargin=17pt,
  framexleftmargin=17pt,
  framexbottommargin=0pt,
  framextopmargin=0pt,
  showstringspaces=false
}

\section{Introduction}
\label{sec:intro}

The stagnating clock speed forced CPU manufacturers into steadily increasing the number of cores on commodity hardware to meet the ever-increasing demand for computational power. Still, the number of parallel processing units on a single GPU is higher by orders of magnitudes. This rich source of computing power became available to general purpose applications as GPUs moved away from single purpose pipelines for graphics processing towards compact clusters of data parallel programmable units~\cite{ohlgs-gc-08}.

Many basic algorithms like sorting and searching, matrix multiplication, or fast Fourier transform have efficient, massively parallel variants.  Also, a large variety of complex algorithms can be fully mapped to the data parallel architecture of GPUs with a massive boost in performance.
Combined with the widespread availability of general-purpose GPU (GPGPU) devices on desktops, laptops and even mobiles, GPGPU computing has been widely recognized as an important optimization strategy. In addition, accelerating coprocessors that better support code branching established on the market.

Since not all tasks can benefit from such specialized devices, developers need to distribute work on the various architectural elements. Managing such a heterogeneous runtime environment inherently increases the complexity. While some loop-based computations can be offloaded to GPUs using OpenACC \cite{c-cnpgo-11} or recent versions of OpenMP \cite{dm-oisas-98} with relatively little programming effort, it has been shown that a consistent task-oriented design exploits the available parallelism more efficiently. Corresponding results achieve better performance~\cite{kwawk-degfg-14} while they are also applicable to more complex work loads. However, manually orchestrating tasks between multiple devices is an error-prone and complex task.

The actor model of computation describes applications in terms of isolated software entities---actors---that communicate by asynchronous message passing. Actors can be distributed across any number of processors or machines by the runtime system as they are not allowed to share state and thus can always be executed in parallel. The message-based decoupling of software entities further enables actors to run on different devices in a heterogeneous environment. Hence, the actor model can simplify software development by hiding the complexity of heterogeneous and distributed deployments.

In this work, we take up our previous contribution \cite{hcs-maeca-15} about actors programmed with OpenCL---the Open Computing Language standardized by the Khronos Group~\cite{khron-15}. We enhance the integration of heterogeneous programming with the C++ Actor Framework \cite{chs-ccafs-14} by a fully effective encapsulation of OpenCL kernels within C++ actors---the OpenCL actor. OpenCL actors can be transparently used at regular actor programming (e.g., like a library or toolset). They can be composed like standard CAF actors, which leads to a multi-staging of OpenCL kernels.  We present indexing as a realistic use case for this elegant programming approach. Furthermore, we thoroughly examine the runtime overhead introduced by our abstraction layer, and justify our integration of heterogeneous hardware to the existing benefits of \caf such as network-transparency, memory-efficiency and high performance.

The remainder of this paper is organized as follows. Section~\ref{sec:background} introduces the actor model as well as heterogeneous computing in general and OpenCL in particular. Our design goals and their realization are laid out in Section~\ref{sec:design}. Along these lines, benefits and limitations of our approach are thoroughly discussed. An application use case that is composed of many data parallel primitives is described in Section~\ref{sec:usecase}. In Section~\ref{sec:evaluation}, we evaluate the performances of our work with a focus on overhead and scalability. Finally, Section~\ref{sec:conclusion} concludes and gives an outlook to future work.

\section{Background and Related Work}
\label{sec:background}

Before showing design details, we first discuss the actor model of computation, heterogeneous computing in general, and OpenCL.

\subsection{The Actor Model}
\label{sec:background:actors}

Actors are concurrent, isolated entities that interact via message passing. They use unique identifiers to address each other transparently in a distributed system. In reaction to a received message, an actor can, (1) send messages to other actors, (2) spawn new actors and (3) change its own behavior to process future messages differently.

These characteristics lead to several advantages. Since actors can only interact via message passing, they never corrupt each others state and thus avoid race conditions by design. Work can be distributed by spawning more actors in a divide and conquer approach. Further, the actor model addresses fault-tolerance in distributed systems by allowing actors to monitor each other. If an actors dies unexpectedly, the runtime system sends a message to each actor monitoring it. This relation can be strengthened through bidirectional monitors called links. By providing network-transparent messaging and fault propagation, the actor model offers a high level of abstraction for application design and development targeted at concurrent and distributed systems.

Hewitt et al.~\cite{hbs-umafa-73} proposed the actor model in 1973 as part of their work on artificial intelligence. Later, Agha formalized the model in his dissertation~\cite{a-amccd-86} and introduced mailboxing for processing actor messages. He created the foundation of an open, external communication~\cite{amst-ttac-92}. At the same time, Armstrong took a more practical approach by developing Erlang~\cite{a-he-07}.

Erlang is a concurrent, dynamically typed programming language developed for programming large-scale, fault-tolerant systems \cite{a-mrdsp-03}. Although Erlang was not build with the actor model in mind, it satisfies its characteristics. New actors, called processes in Erlang, are created by a function called \lstinline^spawn^. Their communication is based on asynchronous message passing. Processes use pattern matching to identify incoming messages.

To combine the benefits of a high level of abstraction and native program execution, we have developed the C++ Actor Framework (\caf) \cite{chs-ccafs-14}.
Actors are implemented as sub-thread entities and run in a cooperative scheduler using work-stealing. As a result, the creation and destruction of actors is a lightweight operation. Uncooperative actors that require access to blocking function calls can be bound to separate threads by the programmer to avoid starvation. Furthermore, \caf includes a runtime inspection tool to help debugging distributed actor systems.

In \caf, actors are created using the function \lstinline^spawn^. It creates actors from either functions or classes and returns a network-transparent actor handle. Communication is based on message passing, using \lstinline^send^ or \lstinline^request^. The latter function expects a response and allows setting a one-shot handler specifically for the response message. This can be done in two ways: either the actor maintains its normal behavior while awaiting the response or it suspends its behavior until the response is received.

\caf offers dynamically as well as statically typed actors. While the dynamic approach is closer to the original actor model, the static approach allows programmers to define a message passing interface which is checked by the compiler for both incoming and outgoing messages.

Messages are buffered at the receiver in order of arrival before they are processed. The behavior of an actor specifies its response to messages it receives. \caf uses partial functions as message handlers, which are implemented using an internal domain-specific language (DSL) for pattern matching. Messages that cannot be matched stay in the buffer until they are discarded manually or handled by another behavior. The behavior can be changed dynamically during message processing.

In previous work~\cite{chs-rapc-16}, we compared \caf to other actor implementations. Namely Erlang, the Java frameworks SALSA Lite~\cite{dv-slhar-14} and ActorFoundry (based on Kilim~\cite{sm-kitaj-08}), the Scala toolkit and runtime Akka~\cite{ti-a-12} and Charm++~\cite{kk-cppwm-96}. We measured (1) actor creation overhead, (2) sending and processing time of message passing implementations, (3) memory consumption for several use cases and (4) picked up a benchmark from the Computer Language Benchmarks Game. The results showed that \caf displays consistent scaling behavior, minimal memory overhead and very high performance.

\subsection{Heterogeneous Computing}
\label{sec:background:heterogeneous}

Graphic processing units (GPUs) were originally developed to calculate high resolution graphic effects in real-time \cite{nd-tgce-10}. High frame rates are achieved by executing a single routine concurrently on many pixels at once. While this is still the dominant use-case, frameworks like OpenCL \cite{s-oahag-11} or CUDA (Compute Unified Device Architecture)~\cite{kh-pmpph-13} offer an API to use the available hardware for non-graphical applications. This approach is called general purpose GPU (GPGPU) computing.

The first graphics cards were build around a pipeline, where each stage offered a different fixed operation with configurable parameters \cite{lkm-upve-01}. Soon, the capabilities supported by the pipeline were neither complex nor general enough to keep up with the developing capabilities of shading and lighting effects. To adapt to the challenges, each pipeline stage evolved to allow individual programmability and include an enhanced instruction set \cite{b-ds-06}. Although this was a major step towards the architecture in use today, the design still lacked mechanisms for load balancing. If one stage required more time than others, the other stages were left idle. Further, the capacities of a stage were fixed and could not be shifted depending on the algorithm. Eventually, the pipelines were replaced by data parallel programmable units to achieve an overall better workload and more flexibility \cite{ohlgs-gc-08}. All units share a memory area for synchronization, while in addition each unit has a local memory area only accessible by its own processing elements. A single unit only supports data parallelism, but a cluster of them can process task parallel algorithms as well.

By now, this architecture can be found in non-GPU hardware as well. Accelerators with the sole purpose of data parallel computing are available on the market. While some have a more similar architecture to GPUs, for example the Nvidia Tesla devices~\cite{n-tcpbb-11}, others are build closer to x86 machines, most prominently the Intel Xeon Phi coprocessors~\cite{i-ixpcp-15}. Both have many more cores than available CPUs and require special programming models to make optimal use of their processing power.

Naturally, algorithms that perform similar work on independent data benefit greatly from the parallelism offered by these architectures. Since most problems cannot be mapped solely to this category, calculations on accelerators are often combined with calculations on the CPU. This combination of several hardware architectures in a single application is called heterogenous computing.

\subsection{\opencl}
\label{sec:background:opencl}

The two major frameworks for GPGPU computing are CUDA (Compute Unified Device Architecture)~\cite{kh-pmpph-13}---a proprietary API by Nvidia---and OpenCL~\cite{s-oahag-11}---a standardized API. In our work, we focus on OpenCL, as it is vendor-independent and allows us to integrate a broad range of hardware. The OpenCL standard is developed by the OpenCL Working Group, a subgroup of the non-profit organization Khronos Group~\cite{khron-15}. Universality is the key feature of OpenCL, but has the downside that it is not possible to exploit all hardware-dependent feature. The OpenCL framework includes an API and a cross-platform programming language called ``\openclc'' \cite{m-tos-12}.

\singlefig{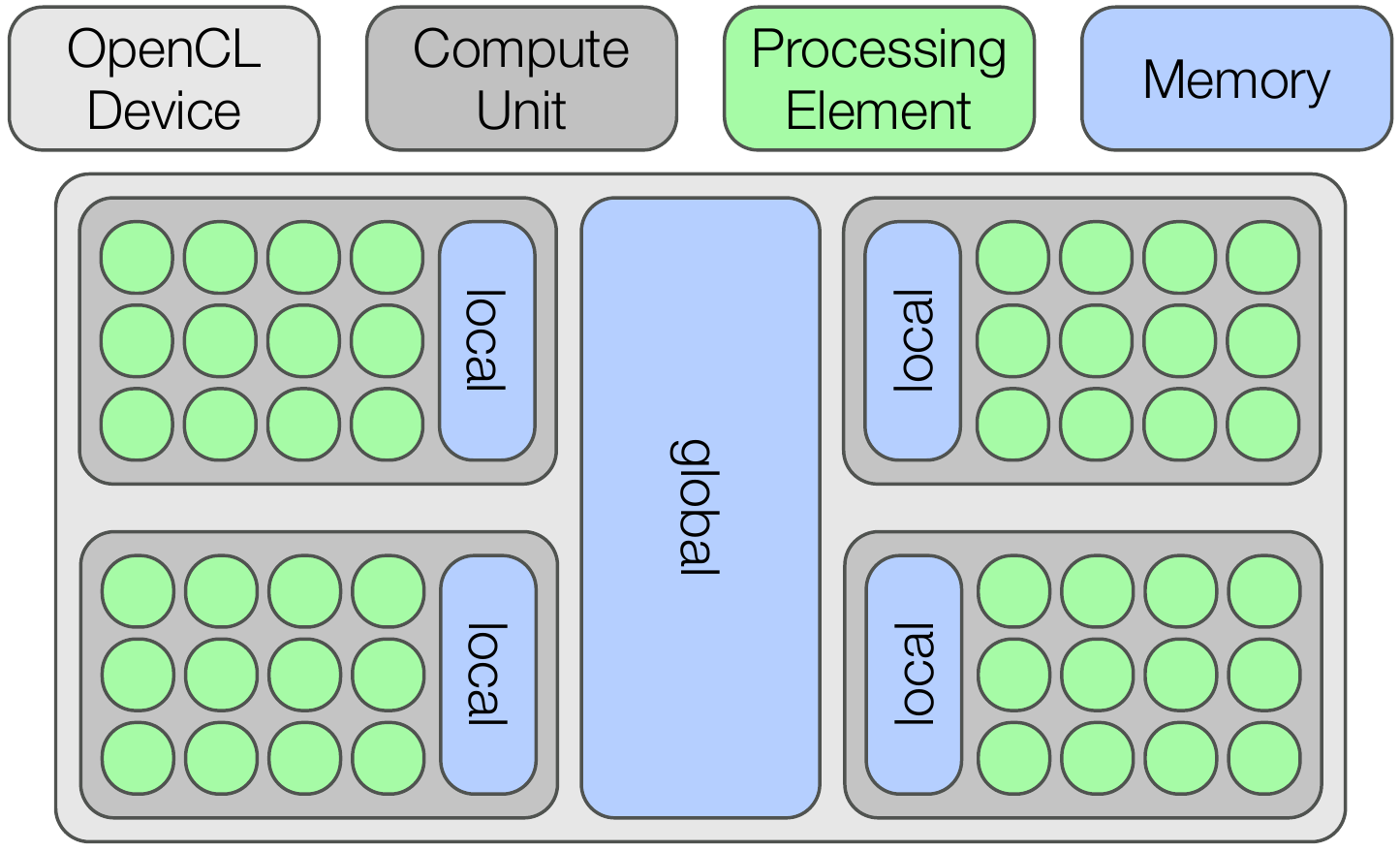}{The OpenCL view on a computation device, dividing the hardware into groups of Processing Elements called Compute Units that have access to shared local memory.}{fig:ocl_device}

A study by Fang et al.~\cite{fvs-cpcco-11} examines the performance differences between OpenCL and CUDA. Their benchmarks are divided into two categories. The first category consists of synthetic benchmarks, which measure peak performance and show similar results for OpenCL and CUDA. The second category includes real-world applications and shows a better overall performance for CUDA. However, the author explain the gap with differences in the programming model, optimizations, architecture and compiler. They continue to define a fair comparison that includes several steps such as individual optimizations and multiple kernel compilation steps.

Figure~\ref{fig:ocl_device} depicts a computing device from the perspective of OpenCL. Each device is divided into compute units (CU), which are further divided into processing elements (PE) that perform the actual calculations. OpenCL defines four different memory regions, which may differ from the physical memory layout. The global memory is accessible by all PEs and has a constant memory region with read-only access. Each local memory region is shared by the PEs of a single CU. In addition, each PE has a private memory region which cannot be accessed by others.

Each OpenCL program consists of two parts. One part runs on the host, normally a CPU, and is called host program. The other part consists of any number of kernels that run on an OpenCL device. A kernel is a function written in an OpenCL-specific C dialect. OpenCL does not require a binary interface, as kernels can be compiled at runtime by the host program for a specific GPGPU device.

A kernel is executed in an $N$-dimensional index space called ``NDRange''. Derived from three dimensional graphic calculations, $N$ can be either one, two or three. Each tuple ($n_{x}$, $n_{y}$, $n_{z}$) in the index space identifies a single kernel execution, called work-item. These tuples are called global IDs and allow the identification of work-items during the kernel execution. Further organization is achieved through work-groups. The number of work-items per work-group cannot exceed the number of processing elements in a compute unit. Similar to the global index space, work-items can be arranged in up to three dimensions inside a work-group. All items in a work-group run in parallel on a single CU. Depending on the available hardware, work-groups may run sequentially or in parallel.

The host program initializes data on the device, compiles kernels, and manages their execution. This requires a series of steps before running a kernel on an OpenCL device. Available device drivers offer an entry point in form of different platforms. These can be queried through the OpenCL API. Once a platform is chosen, all associated device IDs can be acquired. The next step is to create a context object for managing devices of the platform in use.

Communication with a device requires a command queue. The number of command queues per context or device is not limited, though a queue is associated with a single device. Multiple commands can be organized with events. Each command can generate an event which can then be passed to another command to define a dependency between them. Alternatively, OpenCL allows associating an event with a callback. In this way, an asynchronous workflow can be implemented.

Before a kernel---usually stored as source code in the host application---can run on a device, it needs to be compiled using the API of OpenCL. The compilation is then wrapped in a program object. Each program can compile multiple kernels at once and allows their retrieval by name. Running a kernel requires the transfer of its input argument to the target device, as the memory regions of host and GPGPU device are usually disjoint. OpenCL organizes chunks of memory as memory buffer objects that can be created independently and set as read-write, read-only or write-only. Once each argument is assigned to a buffer and the programmer has specified all dimensions in the index space, the kernel can be scheduled. The last step in this process is copying any produced results from the GPGPU device back to the host.

OpenCL offers reference counted handles for its components using the prefix ``\lstinline^cl_^'', e.g., a kernel is stored as a \lstinline^cl_kernel^. The internal reference count needs to be managed manually. In a similar manner, functions are prefixed with \lstinline^cl^, e.g., \lstinline^clGetPlaformIDs^. Most API calls can be executed blocking as well as non-blocking.

The Khronos Group is actively working on advancing OpenCL. The next version of the specification is available as a provisional document since April 2016. In addition to OpenCL itself, the group supports projects that build upon or support OpenCL. SYCL (C++ Single-source Heterogeneous Programming for OpenCL)~\cite{hr-siodm-15} aims to provide the same source code for the CPU and device part, compared to a separate code base for the OpenCL kernels. Since the code for all targets is written with \cpp templates, it can be shared across platforms. However, the specification keeps to the familiar execution model from OpenCL and imposes the same restrictions to the SYCL device code as to \openclc.

\subsection{Nested Parallelism}
\label{sub:background:nested}

The index space (NDRange) specified to execute a kernel can be much larger than the parallelism offered by the hardware. In these cases, processing works by slicing the index space into work groups, which are than executed subsequently or partially in parallel---depending on the hardware characteristics. This concept works well for simple independent calculations. Typical examples that easily scale with the vast parallelism on GPUs are matrix multiplications or calculating fractal images such as the Mandelbrot set. The restrictions of this approach become apparent when moving to more complex applications.

Algorithms that process large amounts of data with dependencies between subsequent computation steps dependent on sequential execution of different kernels. A limiting factor is the number of work items that can be executed in parallel on a device. This number depends on the available compute units and the work items per unit---both vary greatly on different hardware. A problem is the synchronization of processing steps over the total amount of data, which might be much larger than the number of work items that can actually run in parallel. Sorensen et al.~\cite{sdbgr-pibsg-16} discuss these limitations and propose an inter-workgroup barrier that synchronizes all threads running in parallel by estimating the \textit{occupancy}, i.e., the number of work items that can run in parallel, and building a barrier using OpenCL atomics.

The traditional work-flow for complex computations manages this by splitting algorithms into multiple kernel execution stages. Each stage executes all work items before moving to the next stage to ensure that all data has been processes and all work items have a consistent view on the memory. Between each stages the data is kept on the device to prevent expensive copy operations.

Modern GPGPU programming models address this limitation. They allow the creation of new kernels from the device context without additional interaction from the CPU context. This can be seen in the \textit{dynamic parallelism} offered in CUDA or the \textit{nested parallelism} of OpenCL. The OpenCL 2.0 specification introduced the function \lstinline^enqueue_kernel^ for this purpose \cite{mh-tos-15}. It can be used to enqueue new kernels to the command queue from the GPU, specifying a new index space for the execution. Moreover, the enqueued kernel can be run asynchronously, after all work items of the current kernel have finished, or after the work items of the work group have finished. Subsequent kernel executions initiated form the GPU context provide the same guarantees as for subsequent kernel executions form the CPU with regard to synchronization. From the CPU context, nested parallelism looks like a single kernel execution. Thus, waiting for the parent kernel to finish waits for all child kernels as well.

\subsection{Approaches to Heterogeneous Computing}
\label{sub:background:related}

As with multi-core machines, accelerators can be programmed through many different frameworks. The above-mentioned frameworks \opencl and CUDA are the mainstream solutions. They offer specialized functions and the opportunity for optimizations at the price of an extensive API. Many libraries have emerged that use \opencl or CUDA as a back end to offer a higher level API and implementations of often-used algorithms. Examples are Boost.Compute~\footnote{\url{https://github.com/boostorg/compute} (Feb. 2017)} or VexCL~\footnote{\url{https://github.com/ddemidov/vexcl} (Feb. 2017)}.

The projects Aparapi~\cite{a-a-15} and PyOpenCL~\cite{kplci-ppsag-12} provide interfaces for writing OpenCL kernels in their respective language, Java and Python. By avoiding the use of OpenCL C they ease the entrance to heterogeneous computing for developers not familiar with OpenCL. Having this level of abstraction further allows the execution of code on CPUs in case no suitable OpenCL devices is available. While Aparapi provides an interface similar to Java Threads, PyOpenCL relies on annotations to define which functions are offloaded. In contrast, OCCA~\cite{msw-ouaml-14} has the goal to provide portability and flexibility to developers. They contribute a uniform interface for programming OpenMP, CUDA and OpenCL. Writing the offloaded code in macros allows translation depending on the target platform at runtime. An extensible approach allows the addition of new languages in the future.

A pragma-based approach uses code annotations to specify which code should be parallelized by the compiler. A major advantage is the portability of existing code by adding the annotations to the offloaded code blocks. At the same time, the developer has much less control over the execution and less potential for optimization. OpenACC~\cite{os-toapi-13} is a such standard. It supports data parallel computations distributed on many cores as well as vector operations. A comparison between \opencl and OpenACC can be found in the work of Wienke et al.~\cite{wstm-afera-12}. Although \opencl showed much better performance in their tests, the authors conclude that OpenACC opens the field to more programmers and will improve in performance over time.

Opening C++ to GPU programming has been approached from several directions. The C++ framework CuPP \cite{b-cfeci-09} contributes both low-level and high-level interfaces for accessing CUDA to support data parallel applications. With AMP\footnote{\url{http://msdn.microsoft.com/en-us/library/hh265136.aspx} (Feb. 2017)}, Microsoft extends Visual C++ to allow for programming parallel kernels without any knowledge of hardware-specific languages. CAPP \cite{cl-ccabf-15} uses aspects to manage and integrate OpenCL kernels in AspectC++. Also, an interface is provided for executing OpenCL kernels from the C++ code. First steps towards executing OpenCL kernels within the C++ Actor Framework were presented in \cite{cshw-nassp-13}. The basic concept of an OpenCL Actor in CAF \cite{hcs-maeca-15} is now re-examined and extended to execute composable actor chains on resident memory, allowing dynamic execution of data parallel operations. 

Integrating GPU computing into the actor model is also explored by other scientists. For example, Harvey et al.~\cite{hhs-ppaba-15} showed actors running \opencl code as part of the actor based programming language Ensemble. By adding an additional compiler step, they allow the device code to be written in the same language as the rest of their code. This approach simplifies the development as it allows the use of language features such as multi-dimensional arrays. Further optimizations allow the language to keep messages sent between \opencl actors on the device instead of copying it back and forth. The code used as the actors behavior still must be written to address the parallel nature of \opencl devices. Their benchmarks compare OpenACC, Ensemble and native \opencl. In most cases Ensemble performs close to \opencl while OpenACC lacks behind in performance.

\section{The Design of OpenCL Actors}
\label{sec:design}

We are now ready to introduce our approach in detail, discuss its rationales and implementation challenges along with its benefits as well as its limitations.

\subsection{Design Goals and Rationales}
\label{sub:design:goals}

OpenCL is a widely deployed standard containing a programming language (OpenCL C) and a management API. Unlike other approaches such as OCCA~\cite{msw-ouaml-14}, CAF does neither attempt to build a new language unifying CPU and GPGPU programming nor to abstract over multiple GPGPU frameworks. Instead, our approach allows programmers to implement actors using data parallel kernels written in OpenCL C without contributing any boilerplate code. Hence, CAF is hiding the management complexity of OpenCL. 
We want to keep CAF easy to use in practice and confine tools to a standard-compliant C++ compiler with available OpenCL drivers. In particular, we do not require a code generator or compiler extensions.

A possible design option would be to specify a domain-specific language (DSL) for GPGPU programming in C++ based on template expressions. Such a DSL essentially allows a framework to traverse the abstract syntax tree (AST) generated by C++ in order to enable lazy evaluation or to generate output in a different language such as OpenCL C. However, programmers would need to learn this DSL in the same way they  need to learn OpenCL C. Further, we assume GPGPU programmers to have some familiarity or experience with OpenCL or CUDA. Introducing a new language would thus increase the entry barrier instead of lowering it. Also, this would force users to re-write existing OpenCL kernels. For this reason, we chose to support OpenCL C directly.

Our central goals for the design of OpenCL actors are (1) hiding complexity of OpenCL management and (2) seamless integration into CAF with respect to access transparency as well as location transparency.

\paragraph{Hiding Complexity} The OpenCL API is a low-level interface written in \C with a style that does not integrate well with modern \cpp. Although OpenCL does offer a \cpp header that wraps the C API, it shows inconsistencies when handling errors and requires repetitive manual steps. The initialization of OpenCL devices, the compilation and management of kernels as well as the asynchronous events generated by OpenCL can and should be handled by the framework rather than by the programmer. Only relevant decisions shall be left to the user and remain on a much higher level of abstraction than is offered by OpenCL. 

\paragraph{Seamless Integration} OpenCL actors must use the same handle type as actors running on the CPU and implement the same semantics. This is required to make both kinds of actors interchangeable and hide the physical deployment at runtime. Further, using the same handle type enables the runtime to use existing abstraction mechanism for network-transparency, monitoring, and error propagation. Additionally, the API for creating OpenCL actors should follow a conformal design, i.e., the OpenCL abstraction should provide a function that is similar to \lstinline^spawn^.

\subsection{Core Approach to the Integration of OpenCL}
\label{sub:design:core}

The asynchronous API of OpenCL maps well to the asynchronous message passing found in actor systems. For starting a computation, programmers enqueue a task to the command queue of OpenCL and register a callback function that is invoked once the result has been produced. This naturally fits actor messaging, whereas the queue management is done implicitly and a response message is generated instead of relying on user-provided callbacks.

\singlefig{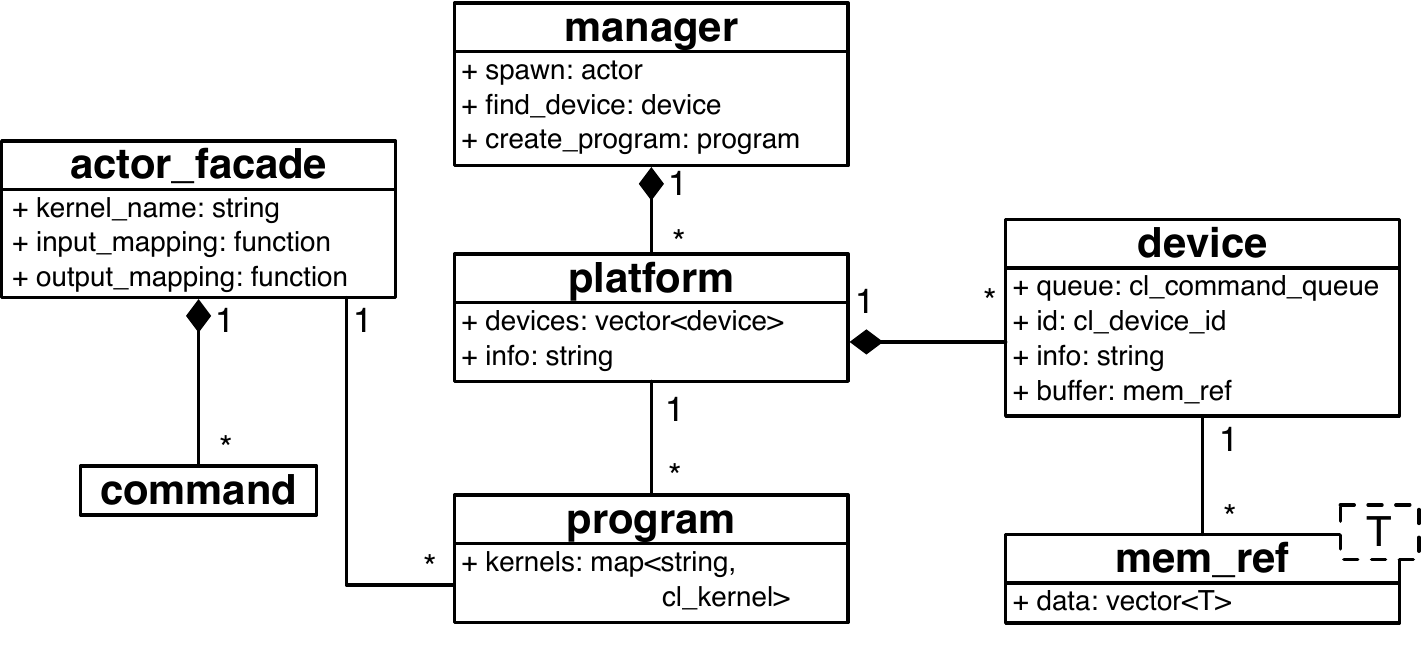}{Class diagram for the OpenCL integration.}{fig:uml}

OpenCL actors introduce easy access to heterogeneous computing within the context of CAF actors. Our main building block is the class \lstinline^actor_facade^ which is shown in Figure~\ref{fig:uml}. The facade wraps the kernel execution on OpenCL devices and provides a message passing interface in form of an actor. For this purpose, the class implements all required interfaces to communicate with other components of CAF (omitted in the diagram for brevity). Whenever a facade receives a message, it creates a \lstinline^command^ which preserves the original context of a message, schedules execution of the kernel and finally produces a result message. The remaining classes implement the bookkeeping required by OpenCL.

\begin{compactitem}
  \item \lstinline^manager^ is a module of an actor system that performs platform discovery lazily on first access and offers an interface to spawn OpenCL actors;
  \item \lstinline^platform^ wraps an OpenCL platform and manages a  \lstinline^cl_context^ that stores OpenCL-internal management data as well as the devices related to the platform;
  \item \lstinline^device^ describes an OpenCL device and provides access to its command queue;
  \item \lstinline^program^ stores compiled OpenCL kernels and provides a mapping from kernel names to objects;
  \item \lstinline^mem_ref^ represents a buffer storing data of type \lstinline^T^ on an OpenCL device and allows its retrieval, both usually handled by the framework.
\end{compactitem}

CAF handles all steps of the OpenCL workflow automatically, but allows for fine-tuning of key aspects. For example, developers can simply provide source code and names for kernels and have CAF create a \lstinline^program^ automatically by selecting a device and compiling the sources. Particularly on host systems with multiple co-processors, programmers may wish to query the \lstinline^manager^ object for accessible devices manually and explicitly create a \lstinline^program^ object by providing a device ID, source code, kernel names, and compiler options.

\subsection{Illustration: Matrix Multiplication with OpenCL Actors}
\label{sub:design:usecase}

We illustrate our concepts and give source code examples for multiplying square matrices. This problem is a straight-forward fit and a common use case for this programming model as each index of the result matrix can be calculated independently. 

\begin{lstlisting}[caption={OpenCL kernel to multiply two square matrices.}, label=lst:kernel]
constexpr const char* name = "m_mult";
constexpr const char* source = R"__(
__kernel void
m_mult(__global float* matrix1,
       __global float* matrix2,
       __global float* output) {
  size_t size = get_global_size(0);
  size_t x = get_global_id(0);
  size_t y = get_global_id(1);
  float result = 0;
  for (size_t idx=0; idx<size; ++idx) {
    result += matrix1[idx + y * size]
            * matrix2[x + idx * size];
  }
  output[x+y*size] = result;
})__";
\end{lstlisting}

Listing~\ref{lst:kernel} shows an OpenCL kernel for multiplying two square matrices stored as string in the variable \lstinline^source^. Additionally, the variable \lstinline^name^ stores the in-source name of the function implementing the kernel. OpenCL requires all kernels to return \lstinline^void^ and use the prefix \lstinline^__kernel^. The first two arguments to the function \lstinline^m_mult^ are two input matrices and the last argument is the result. All matrices are placed in the global memory region to be accessible by all work-items (GPU cores). Since OpenCL does not support multi-dimensional arrays, the matrices are represented as one-dimensional arrays and the position is calculated from the $x$ and $y$ coordinate. At runtime, each instruction will run in parallel on multiple GPU cores but use different memory segments (single instruction, multiple data) identified by the function \lstinline^get_global_id^. In this example, we use two dimensions, which can be queried as index 0 for the $x$ axis and 1 for the $y$ axis. Since we multiply square matrices \lstinline^get_global_size^ returns the same value for both axes.

\subsection{Programming Interface}
\label{sub:design:spawn}

While the OpenCL interface can be translated to actor-like communication in a straightforward way, generating the behavior of the actor is more complex. Since OpenCL source code is compiled at runtime from strings, the C++ compiler needs additional information regarding input and output types.

OpenCL actors are created using a variant of \lstinline^spawn^ that is available through the OpenCL manager when the module is loaded. The execution of a kernel requires configuration parameters like the number of work-items to execute it. Listing~\ref{lst:spawn} illustrates how to create an actor for the kernel shown in Listing~\ref{lst:kernel}. It creates an \lstinline^actor_system^ with a config that loads the module in lines $3-5$.

\begin{lstlisting}[caption={Spawning OpenCL actors.}, label=lst:spawn]
constexpr size_t mx_dim = 1024;
actor_system_config cfg;
cfg.load<opencl::manager>();
actor_system system{cfg};
auto& mngr = system.opencl_manager();
auto worker = mngr.spawn(
  source, name,
  nd_range{dim_vec{mx_dim, mx_dim}},
  in<float>{}, in<float>{}, out<float>{});
auto m = create_matrix(mx_dim * mx_dim);
scoped_actor self;
self->request(worker, m, m).receive(
  [](const vector<float>& result) {
    print_as_matrix(result);
  });
\end{lstlisting}

The first two arguments to the OpenCL \lstinline^spawn^ are strings containing source code and kernel name. CAF will automatically create a \lstinline^program^ object from this source code. For more configuration options, programmers can also create a \lstinline^program^ manually and pass it as the first argument instead. The third argument---the \lstinline^nd_range^---describes the distribution of work-items in three dimensions. A \lstinline^nd_range^ always contains the global dimensions and optionally offsets for the global IDs and local dimensions (to override defaults and fine-tune work-groups in OpenCL). The dimensions are passed as instances of \lstinline^dim_vec^, which is a tuple consisting of either one, two, or three integers. Our example creates one work-item for each index, i.e., $matrix\_size \cdot matrix\_size$ items, meaning that one GPU core computes one element of the result matrix at a time.

The remaining arguments must represent the kernel signature as list of \lstinline^in^, \lstinline^out^, \lstinline^in_out^, \lstinline^local^ and \lstinline^priv^ declarations. This type information allows CAF to automatically generate a pattern for extracting data from messages and to manage OpenCL buffers. While input arguments are provided by the user, storage for output buffers must be allocated by CAF. By default, CAF assumes output buffers to have a size equal to the number of work-items. This default can be overridden by passing a user-defined function to an \lstinline^out^ declaration which calculates the output size depending on the inputs at runtime. The types \lstinline^local^ and \lstinline^priv^ provide buffers in local or private memory for the kernel. In addition to the type of the relate kernel argument, optional template parameters modify the accepted and forwarded types of arguments on the CPU. Specifically, the programmer can choose between values and references to on-device memory. As an example, this enables actors to accept data as a \lstinline^std::vector<T>^ and forward it efficiently to another OpenCL actor as a \lstinline^mem_ref<T>^. 

In our example, the kernel expects two input arguments and one output argument, all represented by one-dimensional dynamic arrays of floating point numbers---in C++ named \lstinline^std::vector<float>^. The template arguments provided here only determine the type of the arguments and thus default to value types (\lstinline^std::vector<T>^) for receipt and answer. In line 13 of Listing~\ref{lst:spawn}, we send two input matrices to the OpenCL actor using \lstinline^request^. The message handler for the result in line 14 awaits the resulting matrix and prints it.

Optionally, programmers can pass two conversion function following the \lstinline^spawn_config^ argument as shown in Listing~\ref{lst:complex}. The first function is then responsible for extracting data from a message while the second function converts the output generated by the kernel to a response message. This mapping gives users full control over the message passing interface of the resulting actor. Per default, these functions are generated by CAF. A message is then matched against all \lstinline^in^ and \lstinline^in_out^ kernel arguments, while the output message is generated from all \lstinline^in_out^ and \lstinline^out^ arguments.

\begin{lstlisting}[caption={Pre- and post-processing in OpenCL actors.}, label=lst:complex]
template <size_t Size>
class square_matrix { /* ... */ };
using fvec = vector<float>;
constexpr size_t mx_dim = 1024;
using mx = square_matrix<mx_dim>;
auto preprocess = [](message& msg)
                  -> optional<message> {
  return msg.apply([](mx& x, mx& y) {
    return make_message(move(x.data()),
                        move(y.data()));
});};
auto postprocess = [] (fvec& res)
                   -> message {
  return make_message(mx{move(res)});
};
auto worker = mngr.spawn(
  kernel_source, kernel_name,
  nd_range{dim_vec{mx_dim, mx_dim}},
  preprocess, postprocess,
  in<float>{}, in<float>{}, out<float>{});
\end{lstlisting}

The example in Listing~\ref{lst:complex} introduces the class \lstinline`square_matrix`, which is used for message passing. Since OpenCL does not allow custom data types, the OpenCL actor needs to convert the matrix to a one-dimensional \lstinline`float` array before copying data to the GPU and do the opposite after receiving the result from OpenCL. This pattern matching step is modeled by the two functions \lstinline`preprocess`, which converts two input matrices to arrays, and \lstinline`postprocess`, which maps a computed array to a matrix.

An OpenCL actor usually sends the result message to the actor that requested the calculation. This behavior can be adapted in multiple ways. First, CAF offers a client side approach for such behavior using the function \lstinline`send_as`. Second, the \lstinline^postprocess^ functions can be used to send messages to other actors using the computed result. Further, automatically sending a response message can be suppressed by returning a default-constructed message. And finally, CAF offers the composition of actor to describe the message flow between a number of actors. This concept will be discussed in more detail in the next section.

\subsection{OpenCL Actors as Kernel Stages}
\label{sub:design:stages}

The basic OpenCL actor offers stateless computation. It transfers the required data between the CPU and the OpenCL device before and after each kernel invocation. This section extends the basic concept by applying OpenCL actors stage-wise and introduces memory objects that can persist over multiple kernel invocations. References to persistent memory are not confined to a specific actor, but represent state available to the execution pipeline. A reference type (see type \lstinline^mem_ref<T>^ depicted in Figure~\ref{fig:uml}) represents data on the GPU device at the CPU, and allows messaging between OpenCL actors to execute subsequent kernels on the same memory. An OpenCL actor that receives such a reference type in a message matches the data type of the reference against the signature of its kernel as it would match incoming data. For this purpose, a reference type includes type information about the data it references in addition to the amount of bytes it refers to and memory access rights. This provides an efficient way to use the output of one kernel as the input for the next kernel.

While each OpenCL stage is a single actor, the composition API of CAF allows to construct a new, composed actor of multiple others. This reduces the need for an additional supervising actor that passes messages from stage to stage. The composition of CAF actors works as follows: Normally, each actor is expected to return its result from a message handler (with \texttt{void} explicitly meaning no result). If a result is not immediately available, though,  actors may return a 'promise' instead.  Such 'promise' indicates that a result will be produced later, or that the result was delegated to another actor which then becomes responsible for responding to the sender. This allows CAF to correlate input with output messages and enables a powerful composition primitive similar to function composition. We denote $C = B \odot A$ to define an actor $C$ which takes any messages it receives as input of $A$ and uses the result as input for $B$. This definition of actors in terms of other actors is intuitively similar to function composition, i.e., $h = f \circ g \equiv h(x) = f(g(x))$.

In the context of OpenCL actors, we need to include the memory transfer to and from the GPU into the actor chain. For this purpose, the first actor in the chain accepts the input data in a message and transfers it to the OpenCL device before forwarding memory references to the next stage. When the references reach the last actor in the pipeline, it reads the results back and sends them to the initial requester, fulfilling the promise. In the composition of $C \odot B \odot A$, the first actor $A$ would transfer the data to the device, perform its computations, and send references to the results to $B$. In turn, $B$ executes its kernel and passes its results to $C$ where the data is read back when $C$ has processed it according to its behavior.

Although the newly composed actor represents a pipeline of kernel executions, it requires messaging between the related CPU actors to pass memory references from one stage to the following. Unless a result from one stage is required to configure the next, kernel executions can be scheduled asynchronously. This allows actors to forward memory references to the next actor before the execution on the device is finished. OpenCL offers an event-based system to express relationships between commands. Here, we use these events to schedule data transfer to devices, followed by a sequence of kernel executions and the final transfer of the results. This allows OpenCL to chain tasks efficiently without downtime for interaction with the CPU.

Listing~\ref{lst:stages:enqueue} depicts how kernels are enqueued into the OpenCL command queue for asynchronous execution. The class \lstinline^command^ is used by an \lstinline^actor_facade^ to wrap a single execution and let it run asynchronously. Only its function \lstinline^enqueue^ is shown here which enqueues the kernel, sets a callback and forwards the arguments to the next actor. It omits error handling for brevity. A reference count manages the lifetime of the command and is incremented when the function is called (line $3$) which represents the reference held by the OpenCL command queue. Next, the command calls the function \lstinline^clEnqueueNDRangeKernel^ to pass the kernel to OpenCL. The kernel arguments were already set by the actor that created the command and require no additional handling here. The first two arguments are the command queue and kernel for the execution, followed configuration of the index space for the execution. This includes the number of dimensions (line $6$), potential offsets for the global index space (line $7$), the global dimensions (line $8$) and the number of work items in a work group (line $9$). The last three arguments enable asynchronous event management in OpenCL: the number of events the kernel has to await, the events themselves and an event that represents the kernel execution itself.

\begin{lstlisting}[caption={Enqueue commands for asynchronous execution.}, label=lst:stages:enqueue]
class command : public ref_counted {
void enqueue() {
  this->ref(); // reference held by the OpenCL command queue
  clEnqueueNDRangeKernel(
    queue, kernel,
    range.dimensions().size(),
    range.offsets(),
    range.dimensions(),
    range.local_dimensions(),
    events.size(), events.data(), &event);
  clSetEventCallback(
    event, CL_COMPLETE,
    [](cl_event, cl_int, void* data) {
      auto cmd = reinterpret_cast<command*>(data);
      cmd->deref();
    },
    this);
  clFlush(queue);
  forward_arguments();
}
};
\end{lstlisting}

After the kernel execution is enqueued, the function \lstinline^clSetEventCallback^ allows registration of a callback when an event is set to a specific execution status. Here, we set a callback to be performed after the event produced by the function \lstinline^clEnqueueNDRangeKernel^ is set to \lstinline^CL_COMPLETE^, i.e., the kernel execution finished. The third argument is the callback itself in form of a C++ lambda. The last argument of the callback is user data that is passed as the last argument when setting the callback, a pointer to the command itself. Inside the callback, this allows decrementing the reference count of the command, thus releasing its resources. After registering the callback, the function \lstinline^clFlush^ ensures that the commands are passed to OpenCL without awaiting their completion. Finally, the function \lstinline^forward_arguments^ sends the memory references expected to be returned by the actor to the next one, bundling the kernel execution event. This enables subsequent kernel stages to schedule their kernels asynchronously to the execution on the device.
The return types of kernel executions can be calculated from the arguments of the spawn function. This allows us to choose a different enqueue implementation when the results of the execution are not limited to reference types. In that case, the callback will only forward the results after the execution has finished and the required data is read back by the CPU from the OpenCL device.

The algorithms performed by stages may differ greatly in complexity, execution time and interface. While the first two characteristics are part of the user implementation without affecting the composition of stages, the last one impacts the compatibility between subsequent stages. Simply passing the output of one kernel to the next may not work. A kernel may produce new data or require local memory for its computations in addition to input arguments and configuration passed along the pipeline. In this case, an incoming message may not contain all arguments required for the execution. For this purpose, stages can create non-input arguments for internal use similar to the basic OpenCL actor. A pre-processing function that is passed to OpenCL actors can add, remove or configure the arguments for the execution while the post-processing function could drop unnecessary output or reorder arguments to fit the next stage. Dropping a reference argument simply releases its memory on the device.

A restriction of pipelining computations is the locality of its kernel execution. A memory reference type is bound to the process memory where it is created as it references a memory object managed by the local OpenCL context. A different process or remote node would have no use of it. This makes pipelines unsuitable for distribution over multiple nodes as is. There are three approaches to handle this: (a) prohibit serialization of the reference type to raise an error when a reference would be sent over the network, (b) introduce automatic memory transfer from the device to the CPU memory before sending the message---this changes the messages content and type---or, (c) include host information of the memory references to allow lazy transfer of the memory should a remote OpenCL actor attempt to use it. The first approach offers the easier solution making expensive copy operations explicit.

While not as efficient as the dynamic parallelism offered by new standards, introducing OpenCL actors as kernel stages raises the expressiveness of CAF and widens its application context to complex GPU-based applications, thereby relying on a widely available version of OpenCL.

\subsection{Design Discussion}
\label{sub:design:discussion}

CAF achieves a much higher level of abstraction than the management API provided by OpenCL. Only key decisions such as the work-item distribution is required by the user. The OpenCL device binding for a kernel defaults to the first discovered device, but can optionally be chosen dynamically at runtime.

The OpenCL actors presented in this section introduce data parallel intra-actor concurrency to CAF. The behavior of an OpenCL actor consists of three parts: (1) a pre-processing function that pattern-matches input messages and forwards extracted data to OpenCL, (2) a data parallel kernel that runs on an OpenCL device, and (3) a post-processing function that finalizes the message processing step and per default converts data produced by the kernel to a response message. Since the data parallel kernel is running in a separate address space and can only use the limited instruction set provided by OpenCL C, sending messages or spawning new actors from OpenCL C directly cannot be achieved. However, the pre- and post-processing functions run on the CPU and allow programmers to spawn more actors and send additional messages in the common way. These two functions can be automatically generated for convenience by deriving all message types from the signature of a kernel.

Transparent message passing and error handling are achieved in our design by mapping the mailbox of an actor to a command queue of OpenCL. From the perspective of the CAF runtime system, an OpenCL actor is not distinguishable from any other actor since it implements the same interfaces as actors running on the CPU. With the \lstinline^spawn^ function of the OpenCL manager, we provide an interface for the creation process of actors that hides most complexity while still granting access to all performance-relevant configuration options via optional parameters.

Once created, the actor handle can be used and addressed independent of its location. The creation process itself has its limitations, though. OpenCL is available for GPUs and dedicated accelerators as well as CPUs. This suggests to run OpenCL actors on the CPU if no other devices are available. While this is conceptually possible, device drivers commonly deployed do not support code compilation for the CPU. Another problem to consider is the workload caused by an OpenCL actor running on the CPU. It is not scheduled with other actors, but competes for the same resources. Alternatively, a single actor could have two implementations, one in OpenCL and one in regular C++. CAF could then choose the implementation that promises the best performance.

At the abstract actor level, the composition of kernel stages offers a way to express dependencies for processing a set of data. Our approach uses actors that wrap a single kernel as building blocks. This design closely follows the idea of the actor model, describing  entities of simple functionality, and is built on top of the composition API available in CAF. It enables a high-level view on kernel pipelines and encourages the reuse of actors for different parts of an algorithm. The downside of this approach is the messaging overhead pass memory references from actor to actor.

An alternative level of composition uses kernels as building blocks to compose a single OpenCL actor that handles multiple kernel stages. This would remove the need for message passing between kernel executions and could prevent idling of the OpenCL device in between kernel executions. To allow such composition a suitable API for handling kernel instances and combining them in an actor would be required. The translation from output of one kernel to the input of the next one could be defined in suitable callbacks similar to the pre- and post-processing functions available for OpenCL actors. This raises the question whether an actor should map to a single kernel---requiring message passing via the CPU---or to an algorithm by wrapping multiple kernel executions---forwarding data by use of callbacks. 

In general, kernel execution and message passing can run in parallel using the event capabilities of OpenCL. Therefore, message passing should only affect execution time if the kernel execution is faster than passing the parameters to the next actor and enqueueing its kernel for execution. The payoff for calculation with heterogeneous hardware rises with the calculation time as the overhead to transfer data between the devices (or pass messages between actors) becomes smaller in comparison. To estimate the costs of message passing between stage actors, we created an actor with an empty kernel and passed it a memory reference to execute its kernel. Measuring the time from sending the message to receiving an answer should give an estimate of the baseline required to process an ``empty'' stage. The measurement results vary between different GPUs and vendors, but mainly remain below 1\,ms. This includes the time required by the OpenCL API and leads us to believe that message passing should not be a bottleneck for most use cases. Looking only at the time between the mapping functions for the output of the first stage and the input mapping function on the second stage is called, the measurements remain around a few microseconds.

With these considerations in mind, an interface that integrates into the actor model and allows for composition based on existing functionality provides flexibility and encourages reuse of kernel stages, whereas an interface for  composing kernels on the OpenCL level allows developers to choose performance over flexibility. A step further towards efficient kernel execution is the nested parallelism discussed in Section~\ref{sub:background:nested}. It wraps multiple kernels into a single actor as the host program cannot differentiate between a simple kernel without child kernels and a kernel enqueueing child kernels from the GPU.

Overall, the introduction of multi-stage actors in addition to the basic OpenCL actor widens the realm of possibilities offered by CAF to developers. It allows to select the right approach for each application: a high level composition of OpenCL actors that fits well into the actor model, or actors that use nested parallelism to implemented kernel stages on the device itself. Implementing an OpenCL actor that wraps kernels on the CPU would provide a intermediate middle option that avoids message passing but still relies on host interaction. Note that nested parallelism depends on the availability of suitable hardware and drivers. While the standard was introduced in 2013, it is not widely available yet and especially lacks support on the hardware deployed.

An advanced aspect of OpenCL usage is scheduling kernels across multiple devices. To enqueue kernels for concurrent execution, a scheduler needs to keep track of the available resources, such as processing elements and memory, as these informations are not offered by OpenCL at runtime. The process get more complicated when using different hardware such as different GPU generations or hardware from various vendors. Depending on the target device, a kernel must be configured specifically for the target to reach optimal performance.

\section{Use Case: Indexing on the GPU}
\label{sec:usecase}

We want to explore the full capabilities of OpenCL actors in CAF by closely following an implementation of a rather complex use case. The creation of bitmap indexes from large volumes of data is a challenging application of sequential kernel executions. Indexing also proved realistic for GPGPU computing, as work from Fusco et al. \cite{fvdd-impps-13} could show. Having VAST~\cite{vps-vupin-16} already as an application domain of CAF, we attempt to accelerate its indexing process through heterogeneous OpenCL computations.

\subsection{Mapping WAH to OpenCL Actors}

Fusco et al. \cite{fvdd-impps-13} presented how WAH~\cite{wos-obiec-06} and PLWAH~\cite{dp-plwah-10} compressions can be created entirely on the GPU. The index consists of blocks of 32 bit values which either represent a heterogeneous sequence of zeros and ones as is---called a 'literal'---or a sequence of homogeneous blocks compressed into a single 'fill'. The corresponding data parallel algorithm consists of six parts applied to the index data. On a high level, it first encodes values with their input position before sorting them by value. This moves values that will be encoded in the same bitmap adjacent to each other while maintaining information about their original distance in the in form of their previous position. From this data, the literals and fills for the WHA encoding are created. The resulting index includes one bitmap for each unique value in the input data. Finally, the algorithm creates a lookup table to find the bitmap related to each value in the index. The six parts are composed from 20 stages which successively run on the GPU without control of the CPU.

For the sake of brevity, we confine our discussion to the step 'fuseFillsLiterals' for building a WAH bitmap index (see Algorithm 5 in~\cite{fvdd-impps-13}). It merges previously computed arrays to build the index. For this purpose, the algorithm interleaves two previously created arrays (chunk ids and literals) and performs a stream compaction on the resulting index. A stream compaction removes all entries from an input array that match a given value, compacting the remaining values. The resulting array should have a length less than or equal to the length of the chunk ids and literals array combined.

Listing~\ref{lst:stages:usecase} presents an actor composition that combines three kernels to perform this algorithm. The first actor prepares the index by merging the chunk ids and literals into the combined index array. Subsequently, the stream compaction removes all zero entries from the index. Billeter et al.~\cite{boa-escws-09} published a stream compaction algorithm for GPUs which is used here. The OpenCL implementation combines phases two and three in a single kernel invocation.

Lines $2$ and $3$ of the listing configure the index spaces for the calculation, see Section~\ref{sub:design:spawn} for details on the configuration parameters. All kernels use a one dimensional index space. While the preparing kernel only requires $k$ work items, each moving a value from the chunk id and literal array into the index, the subsequent stream compaction kernels use one work item per value in the index, i.e., $2 \cdot k$. The stream compaction is written to utilize work-groups of size 128, as declared in the last argument of the \lstinline^range_sc^. Such sizing is applicable and efficient on most GPUs. Next, line $4$ acquires a reference to the OpenCL manager of the local \lstinline^actor_system^.

Three actors are created for this step, one for each kernel (Lines $6$ to $21$). The first actor is created from the program that contains the specific algorithms, in this case the \lstinline^"prepare_index"^ kernel, and is configured to create a one-dimensional index space with $k$ work items. The remaining arguments describe the kernel signature, which requires three \lstinline^uint*^ as input (the first three arguments) and returns two \lstinline^uint*^, the first and last argument. The first argument here is a configuration array passed along the pipeline that contains the number of elements to handle and is used to return newly created values such as the new length after the compaction. The other output is the prepared array for the index.

The kernel signature is not just described by the type of the argument (\lstinline^uint^) but includes tags to specify how the kernel accepts and forwards each argument. As this stage is part of a larger algorithm, the data is forwarded in form of memory references along the pipeline. The \lstinline^in_out^ type requires one parameters for input and one for output while \lstinline^in^ and \lstinline^out^ only require one parameter for the direction they represent.

\begin{lstlisting}[caption={Composing an actor to perform the \textit{fuseFillsLiterals} indexing step.}, label=lst:stages:usecase]
// k is length of the arrays to merge into the index
auto range = nd_range{dim_vec{k}, {}, {}};
auto range_sc = nd_range{dim_vec{2 * k}, {}, dim_vec{128}};
auto& mngr = sys.opencl_manager();

auto prepare = mngr.spawn(
  program_fuse, "prepare_index", range,
  in_out<uint, ref, ref>{},
  in<uint, ref>{}, in<uint, ref>{},
  out<uint, ref>{2 * k});
auto count_elems = mngr.spawn(
  program_sc, "count_elements", range_sc,
  in_out<uint, ref, ref>{}, in_out<uint, ref, ref>{},
  out<uint, ref>{k / 128},
  local<uint>{128});
auto move_elems = mngr.spawn(
  program_sc, "move_valid_elements", range_sc,
  in_out<uint, ref, ref>{},
  in<uint, ref>{}, in<uint, ref>{},
  out<uint, ref>{2 * k},
  local<uint>{128}, local<uint>{128}, local<uint>{128});

// create a composed actor of the three algorithmic steps
auto fuse = move_elems * count_elems * prepare;  
\end{lstlisting}

Next, the actor to handle the first step of the stream compaction is spawned. The stream compaction is located in \lstinline^program_sc^ object which contains the kernels \lstinline^count_elements^ used here and the kernel \lstinline^move_valid_elements^ for the next stage. Arguments kept in local memory can neither be initialized from nor read by the CPU. Moreover, they are not persistent over multiple kernel executions. These buffers are reserved for computations by a work group. Thus, the \lstinline^count_elements^ kernel excepts two arguments in a message: the configuration and the data to compact. It returns an additional argument that is used in the second stage of the stream compaction and contains a value for each work group.

The third actor uses the information calculated by the \lstinline^count_elem^ actor to compact the index, removing all empty entries. For this purpose, it accepts the configuration parameters, the index, and the data calculated in the count stage as input. A new buffer for the compacted index is created by the actor in addition to the three local buffers for each work group. Only the configuration and the new buffer are returned by the actor, which writes the new length of the compacted index into the configuration.

Finally, the three actors are composed into a new actor \lstinline^fuse^, see line $24$. The overall calculation performed by the \lstinline^fuse^ actor can be expressed in the following equation: 
\[
\mbox{\lstinline^fuse^}(msg) = \mbox{\lstinline^move_elems^}( \mbox{\lstinline^count_elems^}( \mbox{\lstinline^prepare^}(msg))). 
\]
It expects the chunk ids and literals computed in an earlier step together with their length as input, and returns the index and its new length as output. The messages sent here between actors only include memory references. The actual data remains on the GPU.

\subsection{Results and Insights}

\singlefig{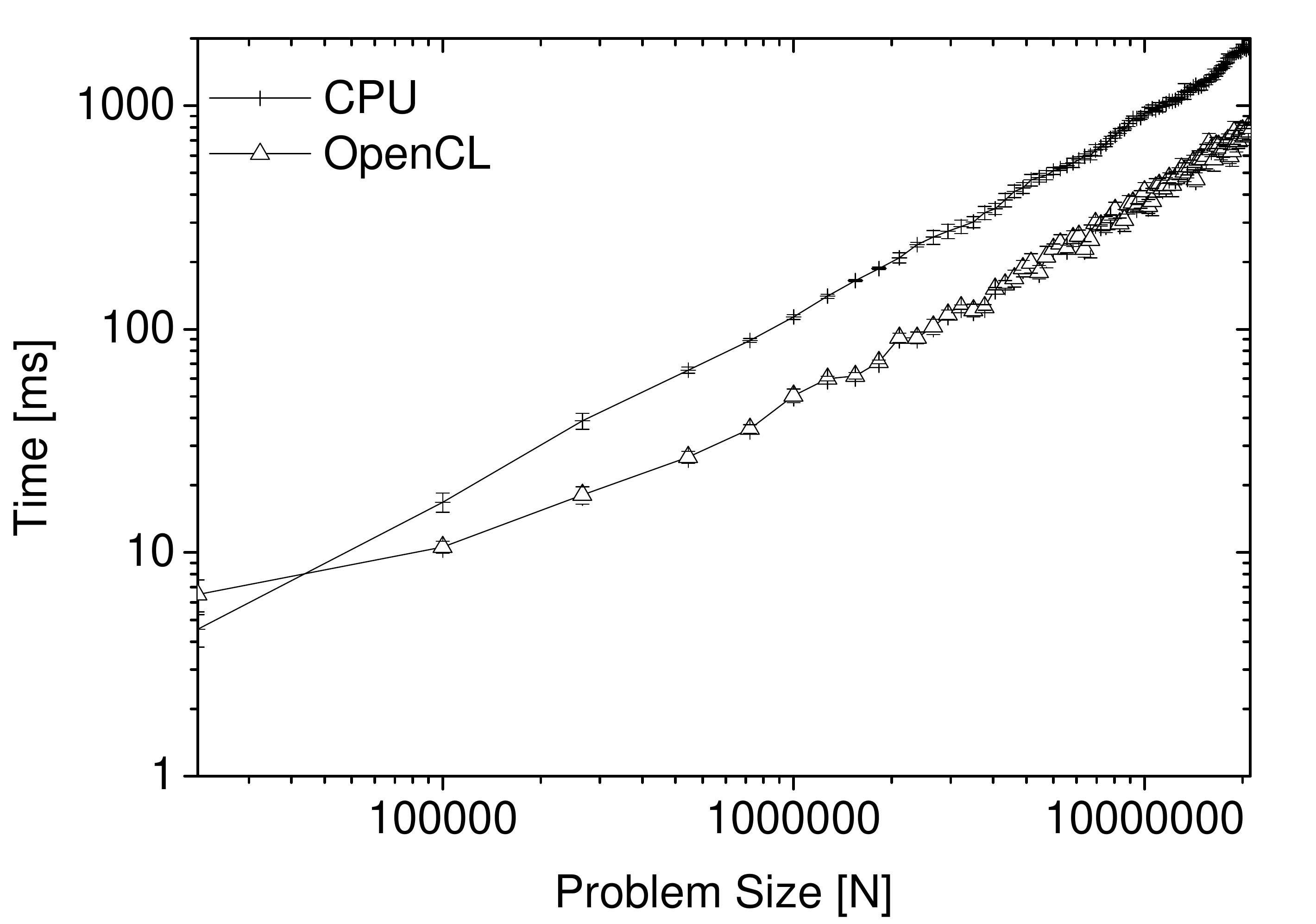}{Runtime for building a WAH index as a function of index size---comparing GPU with CPU performance}{fig:indexing}

We implemented the complete WAH indexing algorithm using \lstinline^libcaf_opencl^. The clean message passing approach provided a familiar environment for writing the algorithm and composing the stages. Standard algorithms such as the stream compaction were repeatedly needed at different stages of the algorithm, offering the opportunity to reuse the respective actors. Radix sort using a fixed cardinality of 16 bits was implemented for ordering the input. All code is publicly available on GitHub.

Going forward, actors that offer such standard algorithms could be included in the framework to provide easier access to building blocks for multi-stage OpenCL actors and GPGPU computing with CAF. The application we developed consisted mostly of OpenCL actors. When the indexing implementation is optimized and proves to be useful for VAST itself, its integration will reveal how well OpenCL actors blend into a larger code base.

While the algorithm is fully functional and produces the expected index, it would still require optimizations to reproduce the absolute performance presented in \cite{fvdd-impps-13}. However, we are rather interested in the qualitative runtime behavior and the scaling.

Figure~\ref{fig:indexing} compares the execution times of creating a WAH bitmap index on the GPU and the CPU as a function of the problem size. Inputs range from $10.000$ to $20.000.000$ values. Initial measurement steps include $N =20.000$, $100.000$, and $250.000$ followed by increments of $250.000$ for each subsequent measurement. Both axes are scaled logarithmically, depicting the means of 10 measurements as well as their standard deviations. We start the timer on the CPU with the initial invocation of indexing, and stop on the CPU after final completion including data return. A Tesla C2075 GPU running OpenCL 1.1 was used for indexing, built in a 24 core Dell server running CentOS 7. The GPU has 14 compute units that can run up to 1024 work items each, adding up to $14.336$ concurrent computations.

The runtime consumption of the CPU grows linearly as expected from the algorithms in use. 
Asymptotically, the GPU also exhibits linear scaling with about half the slope. Correspondingly, execution times on the CPU are about twice as large as on the GPU. For small problem sizes, though, the GPU starts with a slightly higher initialization overhead and a clearly sub-linear growth due to its high inherent parallelism. 

In summary, the results show that indexing can be successfully offloaded to a GPU, leaving the CPU idle for performing regular operations such as interactive searches and queries. GPU offloading grants a significant speedup. Keeping in mind the results of \cite{fvdd-impps-13}, the latter can be expected even higher as the algorithmically complex indexing on the GPU promises higher optimization potentials than the same on the CPU.

\section{Evaluation of the OpenCL Actor}
\label{sec:evaluation}

We have implemented four benchmark programs to systematically measure runtime characteristics and overhead introduced by our OpenCL wrapper.

The first benchmark compares the creation time of OpenCL actors to the event-based actors of CAF. Our next two benchmarks examine the overhead we induce compared to manually using the OpenCL API. Here, we take a look at single calculation before comparing our implementation against an optimized scenario. Our final benchmark examines the scalability in heterogeneous setups by  stepwise transferring workload to a GPU and a coprocessor.

The first benchmarks were performed on a Late 2013 iMac with a 3.5 GHz Intel Core i7 running OS X and OpenCL version 1.2. The GPU is an NVIDIA GeForce GTX 780M GPU with 4096 MB memory. The last  benchmarks on scalability use a machine  with two twelve-core Intel Xeon CPUs clocked at 2.5 GHz equipped with a Tesla C2075 GPU, as well as a Xeon Phi 5110P coprocessor. The server runs Linux and uses the graphics drivers provided by Nvidia (version 375.20) and the Intel OpenCL Runtime 14.2.

\subsection{Spawn Time}
\label{sub:eval:spawn_time}

\singlefig{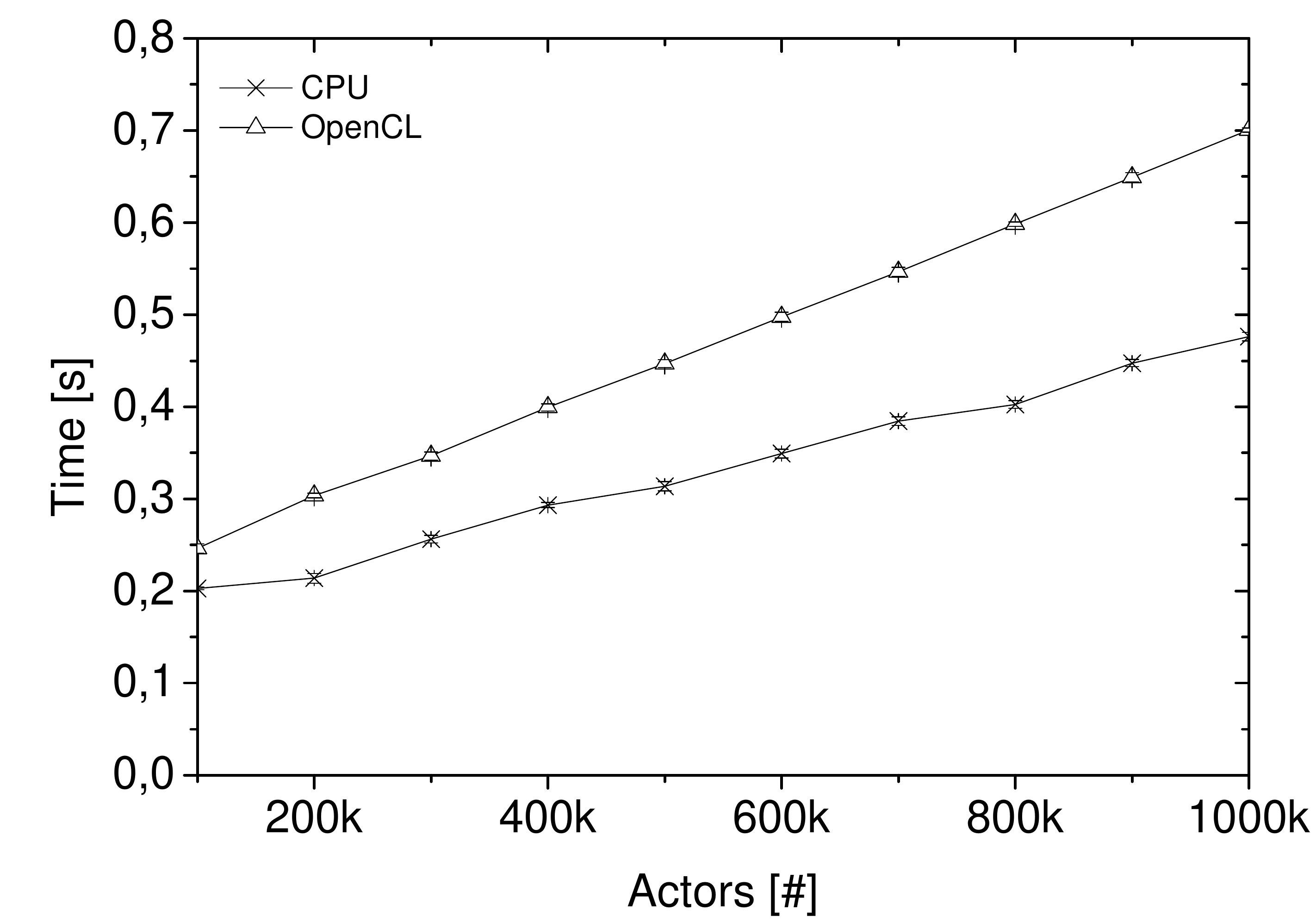}{Comparing the wall-clock time for spawning OpenCL versus event-based actors.}{fig:spawn_time}

Our first benchmark focuses on the time to instantiate OpenCL actors. The creation of actors is traditionally a lightweight operation. We expect the creation of OpenCL actors to be more heavy weight than the creation of other actors in CAF. Still, we want to quantify the overhead associated with actor creation.

We compare the creation time of OpenCL actors to that of event-based actors. Both benchmarks consist of a loop that spawns one actor per iteration. Afterwards we ensure that all actors are active by sending a message to the last created actor and waiting for its response.

The time measured is the wall clock time required to spawn an increasing number of actors. This includes the time required to initialize the runtime environment. To provide an equal setup, we spawn the event-based actors with the \lstinline^lazy_init^ flag. It prevents them from being scheduled for small initialization tasks unless they receive a message, as is the case with OpenCL actors.

Figure~\ref{fig:spawn_time} depicts the wall-clock runtime in seconds as a function of the number of spawned actors. It plots the mean of 50 runs with error bars showing the 95\,\% confidence intervals. In all cases, the error bars are barely visible. Both implementations show a linear dependency with minor growth. However, event-based actors take less time than OpenCL actors and exhibit a smaller slope. The difference in slope indicates a longer  spawn time for each individual OpenCl actor. Similar slopes with a constant distance would have indicated a similar creation time with longer initialization time of the runtime.

Compared to the time required for a simple calculation, the creation time is reasonably small. It is worth mentioning that OpenCL actors are parallelized internally by OpenCL. They are not created as frequently as event-based actors. This limits the overhead further.

\subsection{Runtime Overhead of Actors Over Native OpenCL Programming}
\label{sub:eval:overhead}

\doublefig{Comparing CAF OpenCL Actors with pure OpenCL.}
          {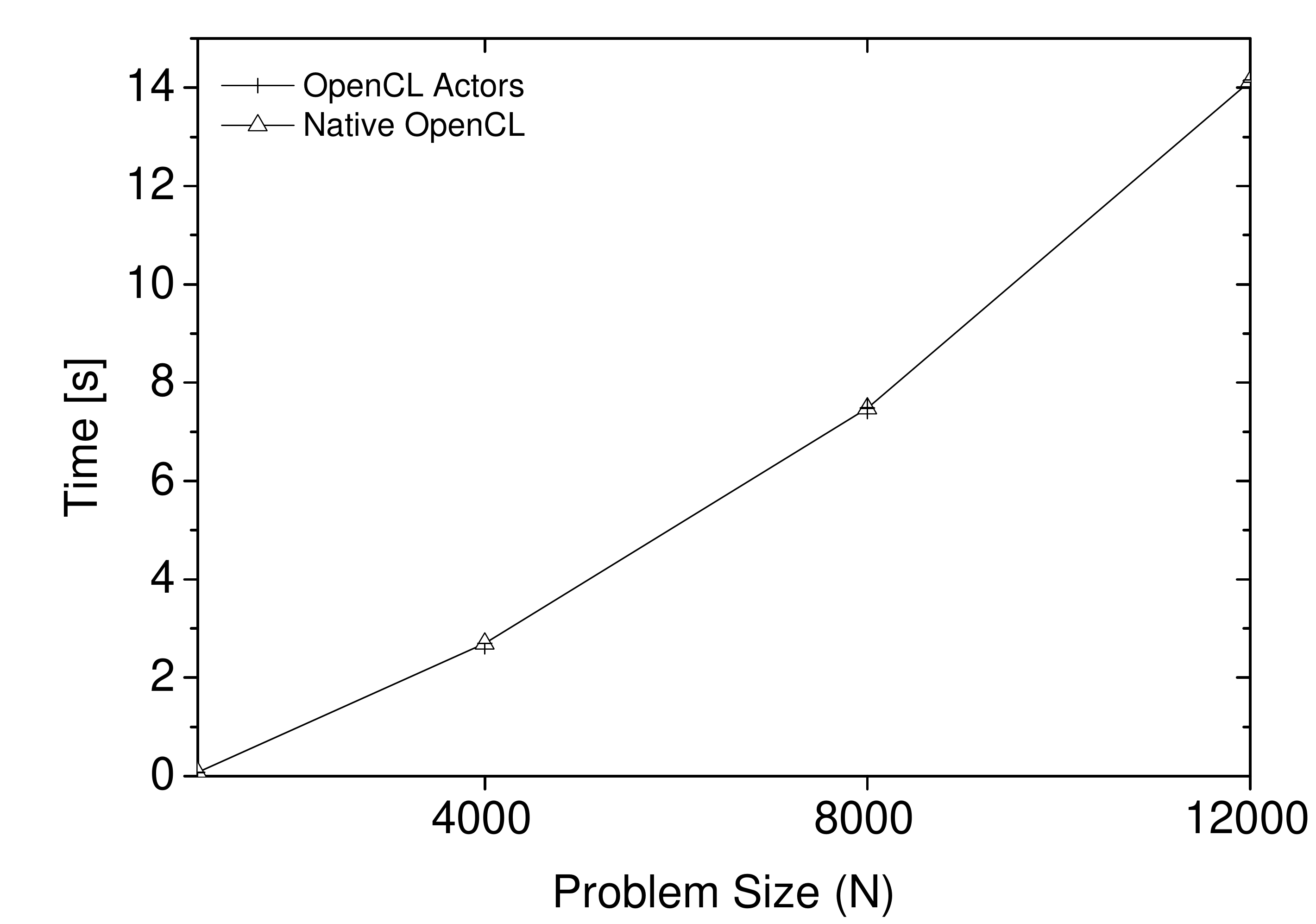}
          {fig:overhead:overhead}
          {The difference between the graphs in (a).}
          {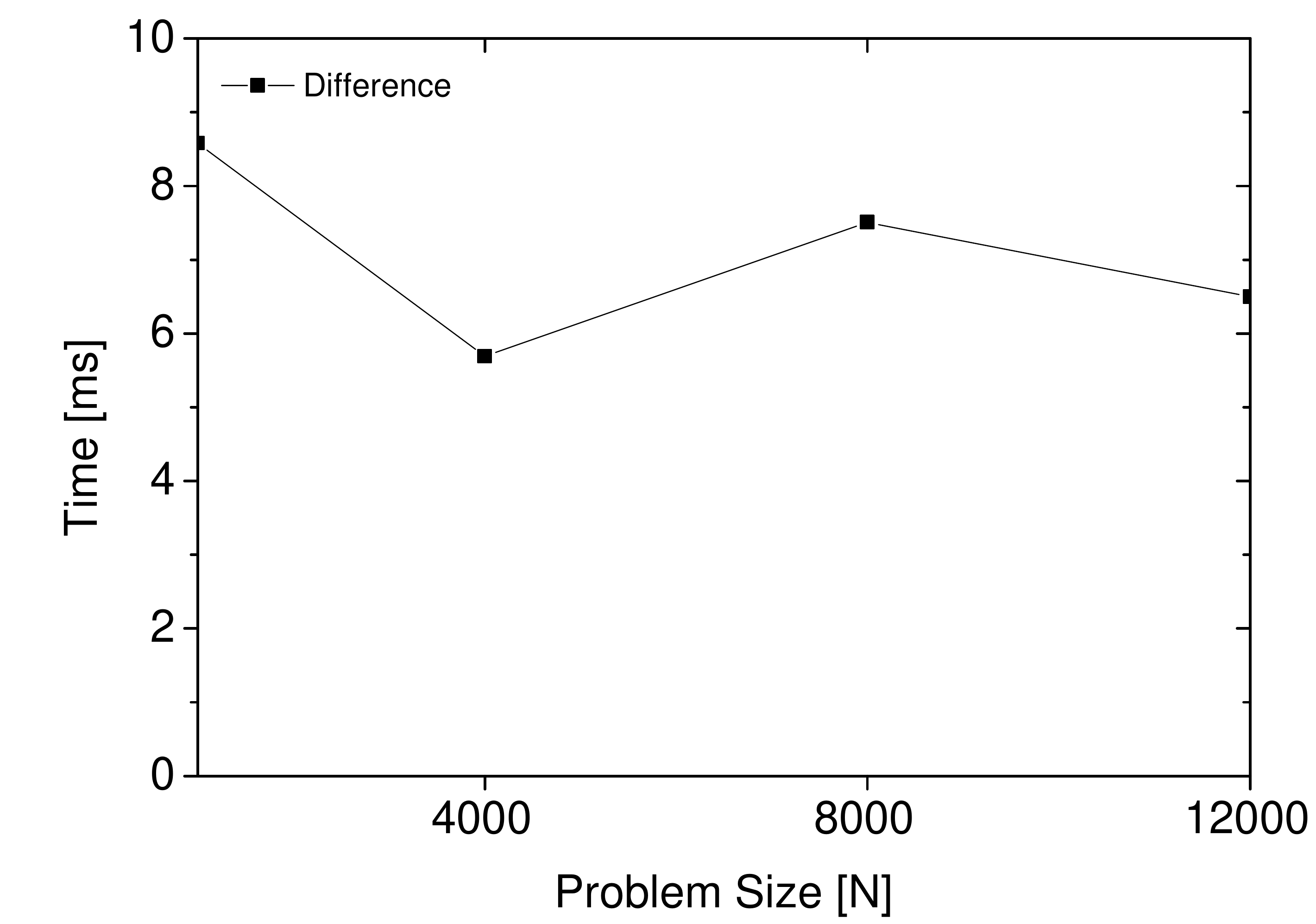}
          {fig:overhead:difference}
          {Overhead of the CAF messaging when multiplying  $N \times N$ matrices.}
          {fig:overhead}

Our second benchmark measures the overhead induced by our actor approach compared to the native API of OpenCL. While the OpenCL actor uses the OpenCL API internally, it performs additional steps such as the setup of the OpenCL environment and the actor creation. This benchmark quantifies the overhead added by message passing and wrapping the OpenCL API.

It implements a program that executes a simple task on a GPU using an OpenCL actor. In this case, the benchmark kernel calculates the product of two $N \times N$ matrices with 1000, 4000, 8000 and 12000 as values for $N$. The increase in problem size shall test for a correlation between the message size and the overhead.

Two measurements are of interest in this case. First, the duration required for the whole calculation, from sending the message to receiving the answer. Second, the time between enqueuing the kernel until OpenCL invokes the callback, which includes data transfer as well as the kernel execution. Ideally, both times should be nearly equal.

Figure~\ref{fig:overhead:overhead} depicts the runtime in seconds as a function of the problem size $N$. Each value is the mean of 50 runs, plotted with the 95\,\% confidence interval. The total calculation time ranges from 0.07\,s up to 14.1\,s. We have also plotted the time difference separately in Figure~\ref{fig:overhead:difference} since the two lines in Figure~\ref{fig:overhead:overhead} are not distinguishable. The difference between the measured values ranges between 5.7\,ms and 8.6\,ms. No discernible slope can be observed in the graph and the measurements fluctuate independently of the problem size.

The results of this measurement clearly show a negligible overhead that does not depend on the problem size. Hence, our high level interface can be used at a very low cost.

\subsection{Baseline Comparison}
\label{sub:eval:baseline}

The previous benchmark examines the overhead for a single calculation by comparing the runtime distribution between CAF and OpenCL. In this benchmark we want to compare the performance when calculating a sequence of independent tasks. Two $1000 \times 1000$ matrices are multiplied with an increasing number of iterations, starting at 1000 and increasing by 1000 in each step up to 10000. The environment is only initialized once and the calculations are preformed sequentially. For CAF, an actor sends a new message when it receives the results of the last calculation. In comparison, the native OpenCL implementation initiates the next calculation as part of the callback. Both programs use the same kernel for the multiplication. We avoid simultaneous kernel executions as we want to examine the overhead in our framework.

\singlefig{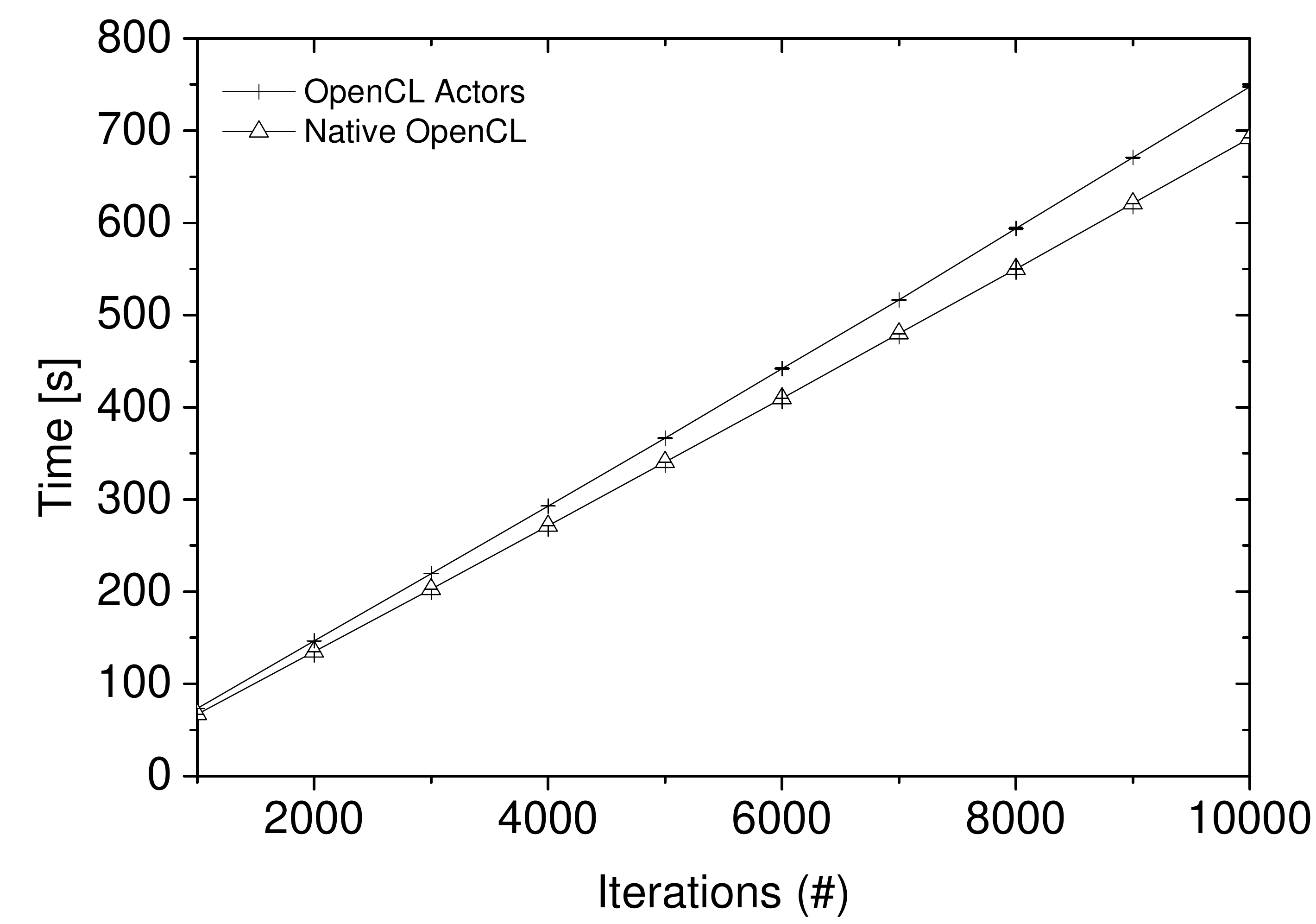}{Comparing the runtime of iterated tasks in CAF versus native OpenCL.}{fig:comparison:comparison}

Figure~\ref{fig:comparison:comparison} displays the wall-clock time as a function of the iterations performed. We plotted the average of 10 measurements as well as a 95\% confidence interval. Since we use the OpenCL API within CAF, it is not possible to achieve a better performance than OpenCL itself. The OpenCL graph is the baseline we aim for with our performance. Both implementations exhibit linear growth. However, the native OpenCL implementation has a smaller slope and the runtime difference between the programs increases. This indicates a consistent overhead required for the message passing compared to the direct API usage. The relative performance difference equals 8.3\,\% for 1000 iterations and slightly decreases to 7.4\,\% at 10000 iterations.

It is worth mentioning that this micro benchmark is looking at a minimal baseline that is not a realistic application scenario. A program using OpenCL will need to include some synchronization to pass GPU-computed results to the CPU and generate the next task for the GPU. Hence, a native application will not meet the baseline simply because it uses the OpenCL API directly.

\subsection{Scaling Behavior in a Heterogeneous Setup}
\label{sub:eval:heterogeneous}

\doublefig{Mandelbrot on the Tesla.}
          {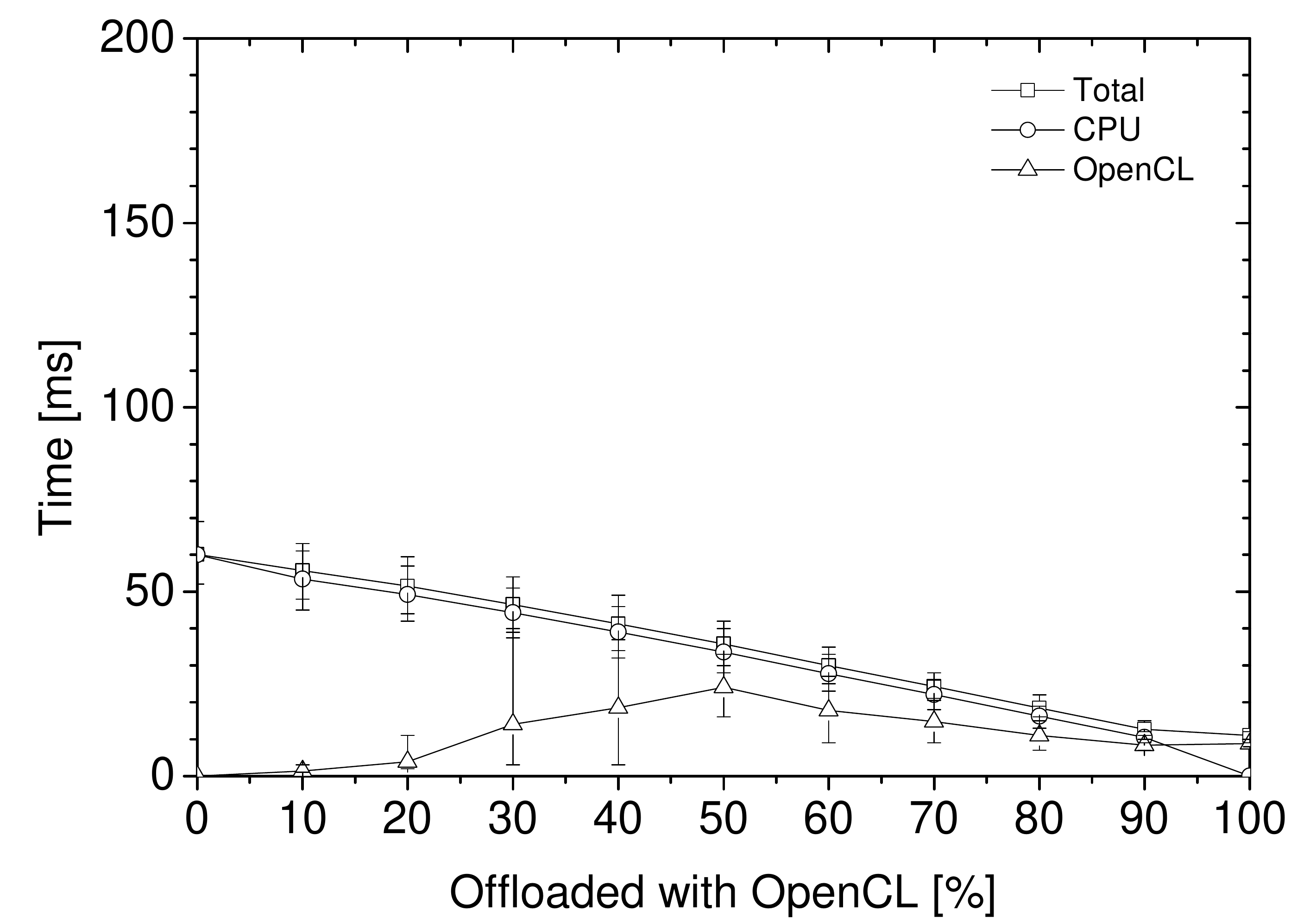}
          {fig:small:tesla}
          {Mandelbrot on the Xeon Phi.}
          {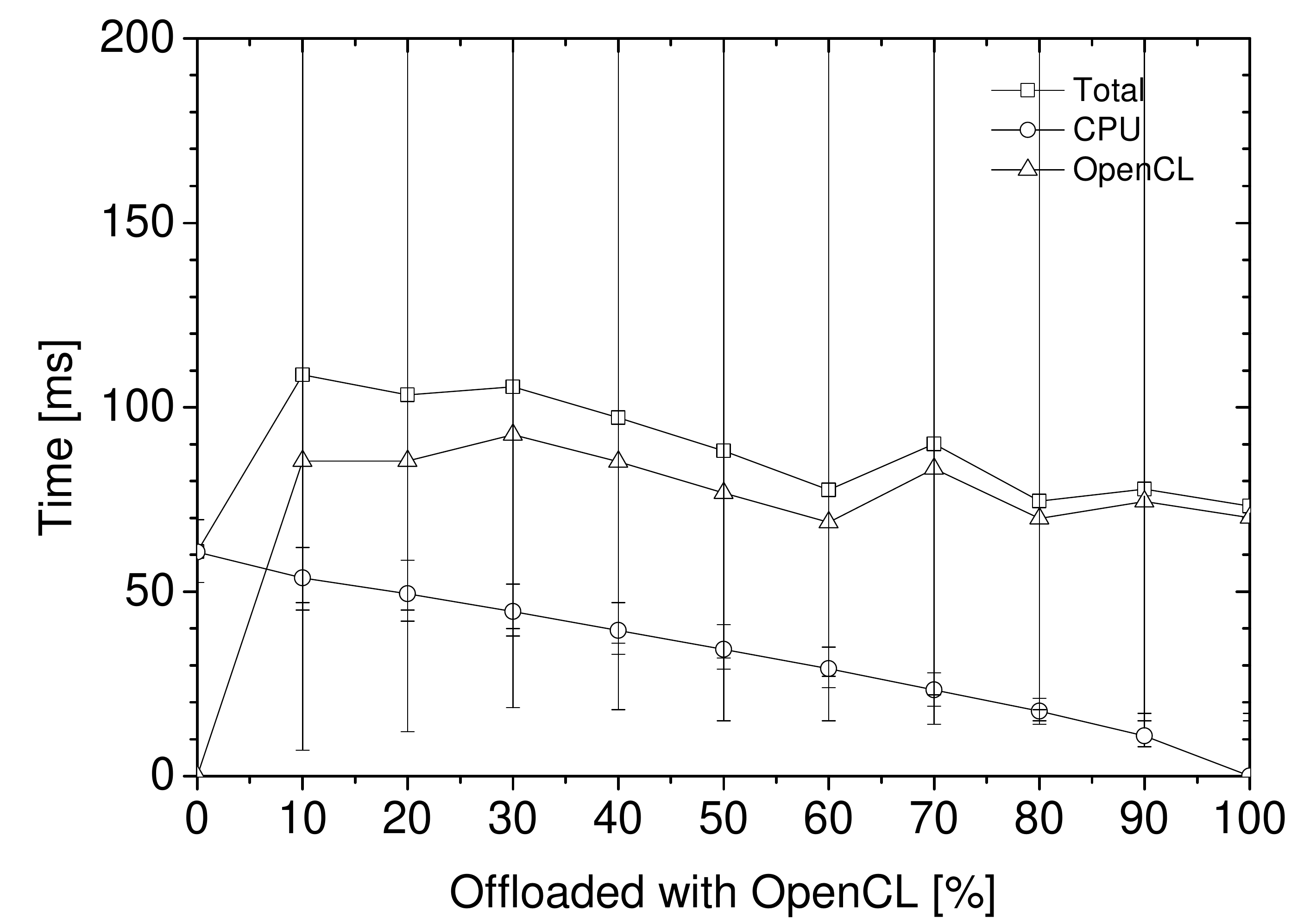}
          {fig:small:phi}
          {Moving a small workload to OpenCL devices.}
          {fig:small}

Our last benchmark focuses on the scalability of our heterogeneous computing approach by incrementally shifting work from the CPU to an OpenCL device. OpenCL distinguishes between CPU, GPU and accelerator devices. Our system includes the two mentioned device, an NVIDIA Tesla GPU and an Intel Xeon Phi accelerator. The difference between a GPU and an accelerator is that GPUs are traditionally used for 3D APIs such as OpenGL or DirectX, while accelerators are dedicated for offloading computations from the host.
The Xeon Phi features an architecture based on x86 processors, although not a compatible one, and differs greatly from the architecture of the Tesla GPU. It consists of 60 cores with 512 bit vector registers and 4 threads each, totaling to up to 240 threads.

We use the calculation of a Mandelbrot set in the benchmark, as the workload can be easily divided into many independent tasks. The problem is a cut from the inner part of a Mandelbrot set that has a balanced processing complexity for the entire image. The workload is offloaded in 11 steps, starting with 0\,\% on the coprocessor and increasing by 10\,\% in each step up to 100\,\%. Each computed image of the Mandelbrot set represents the area of $[-0.5-0.7375i,0.1-0.1375i]$. Our measurements include two different workloads, a resolution of 1920 $\times$ 1080 pixels in Figure~\ref{fig:small} and a resolution of 16000 $\times$ 16000 in Figure~\ref{fig:big}, both measured with 100 iterations. In addition, we increased the number of iterations to 1000 for the larger workload to further examine the scaling behavior.

Figure~\ref{fig:small} depicts the runtime in milliseconds as functions of the problem fraction offloaded. The problem is offloaded to the Tesla in Figure~\ref{fig:small:tesla} and to the Xeon Phi in Figure~\ref{fig:small:phi}. Each graph displays the runtime for the CPU and OpenCL device calculations separately, i.e., the time between starting all actors and their termination. Since calculations are performed in parallel, the total runtime is not a sum of the separate runtimes, but measured independently.

The problem plotted in Figure~\ref{fig:small:tesla} exhibits excellent scalability. The runtime declines until the workload is completely offloaded to the GPU. While the CPU runtime is lower than the total runtime on average, it takes longer to calculate 10~\% of the problem on the CPU than is needed to calculate 100~\% on the GPU. As a result, the lower bound is the time required to process the complete workload on the GPU.

In contrast, Figure~\ref{fig:small:phi} reveals a measurable overhead. While the CPU runtime declines steadily, the runtime measured for OpenCL fluctuates heavily and the total execution time doubles when offloading 10\,\% of work to the Phi. Even when running 100\,\% of the problem size on the Phi, the computation is still slower than the initially measured 60 milliseconds for the CPU-only setup. The initial cost of offloading computations to the Phi are not amortized by faster, parallel computations on the accelerator device. It is worth mentioning that we did not optimize the OpenCL kernel for the Phi, which might result in suboptimal performance on this device.

In summary, these experiments reveal excellent scalability of programming GPUs with CAF actors. However, offloading work to the Xeon Phi is not advisable for this problem size. Since the performance of OpenCL applications largely depends on the driver implementation and configuration, it is left to future work to examine the Phi results in more detail.

\doublefig{Calculation with a 100 iterations.}
          {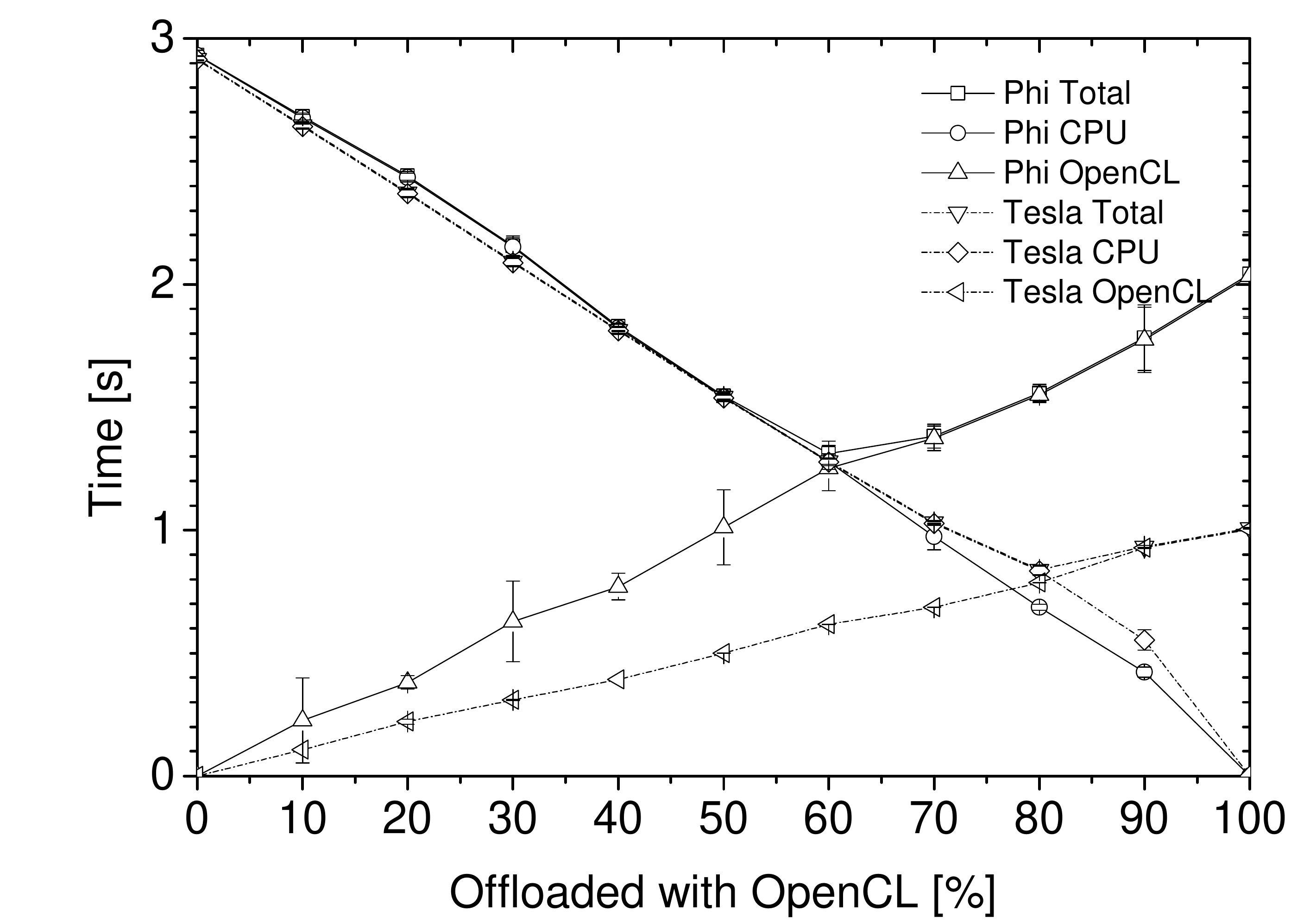}
          {fig:big:small}
          {Calculation with a 1000 iterations.}
          {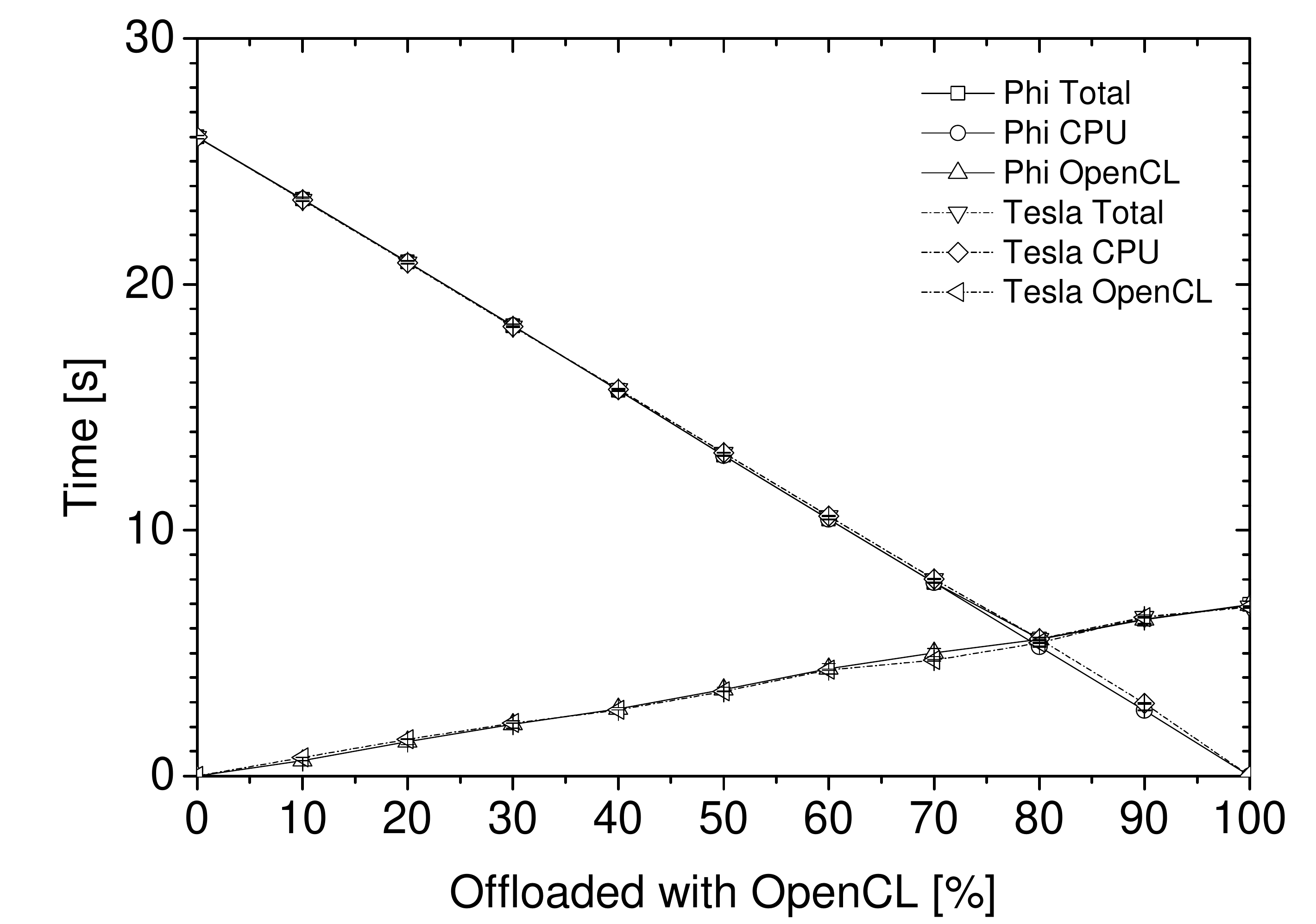}
          {fig:big:big}
          {Moving large workloads to OpenCL devices.}
          {fig:big}

Figure~\ref{fig:big} shows the runtime in milliseconds as a function of the offloaded problem in \% for a larger Mandelbrot image. We have increased the number of pixels from 1920\,$\times$\,1080 to 16000\,$\times$\,16000. The larger image drastically increases the computation time on the device to offset the initial cost of offloading computations. We have run the benchmark using 100 and 1000 iterations per pixel.

Figure~\ref{fig:big:small} visualizes the smaller measurements with 100 iterations for both the Tesla and the Xeon Phi. In difference to the previous benchmark in Figure~\ref{fig:small:tesla}, the best performance is achieved at around 80\,\% on the GPU and around 60\,\% on the accelerator. Since the initial cost of offloading the computation is smaller in comparison to the overall runtime, the Xeon Phi achieves drastically better performance as shown in Figure~\ref{fig:small:phi}, but does not reach the performance of the Tesla.

Finally, the measurements with 1000 iterations are depicted in Figure~\ref{fig:big:big}. Here, the Phi and Tesla perform equally well. Since this setup has the same data rate as before but an increased runtime on the device, it becomes evident that the data transport to the Phi did hinder better results in the previous benchmarks. Hence, this accelerator (with current drivers) is best suited for problems of small data size but large computation demands. 

In a naive approach, we simply transferred a problem from the Tesla to the Phi. This proved to be inefficient for small problems, but improved with an increase in problem size. As should be noted again, optimizing kernels and configurations for the Phi may improve its performance for smaller problems.

\section{Conclusions and Outlook}
\label{sec:conclusion}

Integrating GPGPU computing into an application can increase its performance by orders of magnitudes while releasing the CPU. This holds on all scales from mobiles to server systems. The challenge of integrating GPGPU devices into applications, though, is left to a programmer, who is faced with an ever-growing complexity of hardware architectures and APIs.

The actor model is an important concept for taming the complexity of parallel and concurrent systems and the task-oriented work flow of actors fits the work flow of GPGPU computing very well. The present work on OpenCL actors within the C++ Actor Framework (CAF) shows that an intelligent actor runtime can manage GPGPU devices autonomously while inducing minimal performance overhead. We  optimized performance by designing OpenCL actors as composable stages that pipeline data flows and avoid costly memory transfer between kernel invocations. Supporting OpenCL as first-class implementation option in CAF further broadens the scope of our native actor system by introducing data parallel intra-actor concurrency.

Our presented implementation of OpenCL actors is based on OpenCL 1.1. This version is available across Intel, NVIDIA, and AMD drivers. Our directions for future development fall into three categories:
(1) improve scheduling by load balancing across multiple OpenCL devices both locally and in a network,
(2) extend the use case of indexing on GPUs to common indexing algorithms such as PLWAH,
and (3) provide efficient algorithmic primitives as building blocks for OpenCL actor pipelines.

\subsection*{A Note on Reproducibility}

We explicitly support reproducible research~\cite{acmrep,swgsc-terrc-17}.
Our experiments have been conducted in a transparent standard environment.
The source code of our implementations (including scripts to setup the experiments, CAF measurement apps etc.) are available on GitHub at \url{https://github.com/inetrg/Agere-LNCS-2017}.

\section*{Acknowledgments}

The authors would like to thank Marian Triebe and Sebastian Bartels for implementing benchmarks, testing, and bugfixing.
We further want to thank  Matthias Vallentin for raising the indexing use case, and the iNET working group for vivid discussions and inspiring suggestions.
Funding by the German Federal Ministry of Education and Research within the projects ScaleCast and X--CHECK is gratefully acknowledged.

\bibliographystyle{splncs03}
\balance
\bibliography{own,programming,rfcs,security,gpgpu,meta}

\end{document}